\documentclass[twocolumn,aps,prd,nofootinbib,superscriptaddress,tightenlines]{revtex4}

\usepackage{amsmath, amsfonts, amsthm, amssymb, graphicx, color, hyperref, subfigure}

\allowdisplaybreaks[3]

\newcommand{\bwt} {\begin{widetext}}
\newcommand{\ewt} {\end{widetext}}
\newcommand{\be} {\begin{equation}}
\newcommand{\ee} {\end{equation}}
\def \bal#1\eal  {\begin{align} #1 \end{align}}
\newcommand{\nn} {\nonumber\\}

\newcommand{\ud} {\mathrm{d}}
\newcommand{\uD} {\mathrm{D}}
\newcommand{\dd} {\delta}
\newcommand{\nd} {\nabla}
\newcommand{\pd} {\partial}

\newcommand{\mc} {\mathcal}
\newcommand{\mn} {{\mu\nu}}
\newcommand{\rs} {{\rho\sigma}}

\begin{document}

\title{Black hole hair in generalized scalar-tensor gravity: An explicit example}

\author{Thomas P.~Sotiriou}
\email[]{thomas.sotiriou@nottingham.ac.uk}
\affiliation{School of Mathematical Sciences \& School of Physics and Astronomy, University of Nottingham, Nottingham, NG7 2RD, UK}

\author{Shuang-Yong Zhou}
\email[]{szhou@sissa.it}
\affiliation{SISSA, Via Bonomea 265, 34136, Trieste, Italy and INFN, Sezione di Trieste, Italy}

\date{\today}

\begin{abstract}

In a recent Letter we have shown that in shift-symmetric Horndeski theory the scalar field is forced to obtain a nontrivial configuration in black hole spacetimes, unless a linear coupling with the Gauss-Bonnet invariant is tuned away.  As a result, black holes generically have hair in this theory. In this companion paper, we first review our argument and discuss it in more detail. We then present actual black hole solutions in the simplest case of a theory with the linear scalar-Gauss--Bonnet coupling. We generate exact solutions numerically for a wide range of values of the coupling and also construct analytic solutions perturbatively in the small coupling limit. Comparison of the two types of solutions indicates that non-linear effects that are not captured by the perturbative solution lead to a finite area, as opposed to a central, singularity. Remarkably,
 black holes have a minimum size, controlled by the length scale associated with the scalar-Gauss--Bonnet coupling. We also compute some phenomenological observables for the numerical solution for a wide range of values of the scalar-Gauss--Bonnet coupling.  Deviations from the Schwarzschild geometry are generically very small.
\end{abstract}

\maketitle

\section{Introduction}

In terms of the number of physical parameters one needs to fully characterize them, black holes in gravity theories are rather simple. After the discovery of the uniqueness theorems in general relativity for the Schwarzschild, Reissner-Nordstrom and Kerr solutions \cite{israel1, israel2, Carter:1971zc, Wald:1971iw}, it was conjectured that ``black holes have no hair'' \cite{wheeler} other than the mass, angular momentum and electromagnetic charge. Hawking further proved that stationary black holes, which are the endpoint of gravitational collapse, in  Einstein-Maxwell theory must be of the Kerr-Newman form \cite{Hawking:1971vc}, and later extended this result to Brans-Dicke theory \cite{Hawking:1972qk}. Standard scalar-tensor theories
have also been shown to fall in the reign of the no-hair theorems \cite{Bekenstein:1995un, Sotiriou:2011dz}. Note that the no-hair theorems concern mostly black hole solutions with flat asymptotics, and hairy black holes with non-flat asymptotics often exist. The no-hair theorems inspired the development of black hole thermodynamics. However, it has been realized that hairy black holes do exist when Yang-Mills fields \cite{Volkov:1989fi, Bizon:1990sr, Greene:1992fw} and non-canonical scalars such as Skyrmions \cite{Luckock:1986tr, Droz:1991cx} are included, or when symmetries for non-gravitational fields are relaxed \cite{Herdeiro:2014goa}. The possibility of having black holes with quantum hair has been discussed in Ref.~\cite{Coleman:1991ku}.

Various non-canonical scalar fields ({\em i.e.}~beyond standard scalar-tensor theory) have been employed to build inflation models in the early universe and dark energy models in the late universe. Like Skyrmions, these fields can arise as effective field theories from some underlying theories. Galileon fields \cite{Nicolis:2008in} are one of this kind and have been extensively investigated recently. The prototype of a galileon comes from the Dvali-Gabadadze-Porrati braneworld model \cite{Dvali:2000hr}, where the brane bending mode roughly plays the role of the galileon field \cite{Luty:2003vm}. One central feature of galileon theory is that its Lagrangian terms may contain more than 2 derivatives but its equations of motion remain second order, therefore avoiding Ostrogradski ghosts. This delicate construction makes use of the galileon symmetry $\phi\to \phi+c +b_\mu x^\mu$ ($c,b_\mu$ being constant) \cite{Nicolis:2008in} that becomes exact in the flat space limit. Although the galileon Lagrangian terms can be derived from a braneworld setup \cite{deRham:2010eu}, it is often more convenient to use the four-dimensional covariant effective theories as a proxy (see e.g.~\cite{Deffayet:2009wt, Deffayet:2009mn, Zhou:2011ix}). In generalized galileon theories, the galileon symmetry in the flat space limit is abandoned in order to get the most general second order scalar-tensor theory \cite{Deffayet:2011gz}. Generalized galileon has been shown to be equivalent to Horndeski's theory \cite{Horndeski:1974wa} in four dimensions \cite{Kobayashi:2011nu}. See Ref.~\cite{Deffayet:2013lga} for a recent review of galileon gravity. As they are much more complicated than standard scalar-tensor theory, generalized galileon theories fall outside the purview of the already known black hole no-hair theorems.

Recently, a no-hair theorem for static, spherically symmetric, asymptotically flat black holes has been proposed for the most general scalar-tensor theory with a shift-symmetric scalar that leads to second order equations \cite{Hui:2012qt}. The basic steps of the proof laid out in Ref.~\cite{Hui:2012qt} are the following. Shift symmetry implies that one may write the scalar equation of motion as a current conservation equation $\nd_\mu J^\mu =0$, $J^\mu$ being the associated Noether current. Because of staticity and spherical symmetry, the only non-vanishing component of the current is $J^r$ and is a function of the radial coordinate $r$ only. It is then shown that asymptotic flatness combined with the requirement that the scalar $J^\mu J_\mu$ be regular on the horizon implies $J^r=0$ at all $r$. The last step is to prove that $J^r=0$ at all $r$ necessarily implies $\phi={\rm const}$. It is argued that the current takes the form $J^r = g^{rr}\phi'F(\phi';g,g',g'')$ for general shift-symmetric scalar-tensor theory featuring a canonical kinetic term, where a prime means $\ud/\ud r$, $g$ represents the metric components and $F$ asymptotes a nonzero constant when $\phi'$ goes to zero and $g^{rr}$ goes to 1 near spatial infinity. If this is valid, the only solution for $J^r $ to remain zero at all $r$ is that $\phi'=0$ at all $r$, which means there is no scalar hair for any spherically symmetric black holes. 

In Ref.~\cite{Sotiriou:2013qea} we pointed out that the last step of the proof of Ref.~\cite{Hui:2012qt} discussed above has a loophole. In this paper we will elaborate further on this and also provide a counter example by constructing an explicit hairy black hole solution in a simple model that fall within generalized galileon or Horndeski theories.

The rest of the paper is organized as follows: In the next Section, we first provide a brief overview of Horndeski theories and we rigorously identify the most general shift symmetric generalized galileon (SSGG). We then discuss the form of the current associate with shift symmetry and we carefully analyze the potential loopholes in the proof of Ref.~\cite{Hui:2012qt}. We also explain why a generic SSGG would actually not be covered by this no-hair theorem without fine-tuning or additional symmetry assumptions. In Section \ref{sec:countexam} we present actual black hole solutions for the simplest  theory that evades the no-hair theorem. We first generate exact solutions numerically and we then proceed to construct explicit, analytic solutions perturbatively, in the small-coupling approximation. The structure of these spacetime is fully analyzed and a comparison of the two types of solutions identifies the regime in which  non-linear effects are crucial. Finally, we turn our attention to phenomenology and assess how much our solutions deviate from the Schwarzschild solution. Section \ref{sec:concl} contains our conclusions.

\section{Black hole hair in galileon gravity}
\label{sec:ghbh}

As mentioned in the introduction, we will look more carefully at the last step of the proof of \cite{Hui:2012qt}. In particular, we will examine in detail the functional form of the current associated with shift symmetry and whether it indeed always agrees with what was assumed in Ref.~\cite{Hui:2012qt}. Since we are interested in second order theories (which avoid Ostrogradski instabilities), we can focus on the shift-symmetric restriction of generalized galileon theories \cite{Deffayet:2011gz} (or Horndeski's theory \cite{Horndeski:1974wa}). 

\subsection{The shift symmetry current}

First, we need to identify the most general shift-symmetric generalized galileon action (SSGG). This can be done by imposing shift symmetry on the equations of motion of generalized galileon. We present a detailed discussion about this in Appendix \ref{sec:ssgg} and only give the final result here. The Lagrangian is
\bal
\label{ssgg}
\mc{L} &= \mc{L}_2+\mc{L}_3+\mc{L}_4+\mc{L}_5 ,
\\
\label{ssggl2}
\mc{L}_2 &= K(X) ,
\\ 
\mc{L}_3 &= -G_3(X) \Box \phi ,
\\
\mc{L}_4 &= G_4(X) R + G_{4X} \left[ (\Box \phi)^2 -(\nabla_\mu\nabla_\nu\phi)^2 \right] ,
\\
\label{ssggl5}
\mc{L}_5 &= G_5(X) G_{\mu\nu}\nabla^\mu \nabla^\nu \phi - \frac{1}{6} G_{5X} \big[ (\Box \phi)^3
\nn
&~~~  - 3\Box \phi(\nabla_\mu\nabla_\nu\phi)^2 + 2(\nabla_\mu\nabla_\nu\phi)^3 \big] ,
\eal
where $K, G_3, G_4, G_5$ are arbitrary functions of $X=- \pd^\mu \phi \pd_\mu \phi/2$, $f_X$ means $\pd f(X)/\pd X$, $G_{\mu\nu}$ is the Einstein tensor, $(\nabla_\mu\nabla_\nu\phi)^2\equiv \nabla_\mu\nabla_\nu\phi \nabla^\nu\nabla^\mu\phi$ and $(\nabla_\mu\nabla_\nu\phi)^3=\nabla_\mu\nabla_\nu\phi \nabla^\nu\nabla^\rho\phi \nabla_\rho\nabla^\mu\phi$. The Noether current associated with $\phi\to \phi +\epsilon$ in SSGG is given by
\bal
J^\mu 
 &= -\pd^\mu\phi \bigg( K_X - G_{3X} \Box \phi +G_{4X} R 
 \nn
 &~~~~ + G_{4XX} \left[ (\Box \phi)^2 -(\nabla_\rho\nabla_\sigma\phi)^2 \right] +G_{5X} G^{\rho\sigma}\nabla_{\rho}\nabla_{\sigma}\phi  
 \nn
&~~~~  -\frac{G_{5XX}}{6} \left[ (\Box \phi)^3 - 3\Box \phi(\nabla_\rho\nabla_\sigma\phi)^2 + 2(\nabla_\rho\nabla_\sigma\phi)^3 \right] \bigg)
\nn
&~~~~ -\pd^\nu X \bigg( - \delta^\mu_\nu G_{3X} + 2 G_{4XX} (\Box \phi \delta^\mu_\nu - \nabla^\mu\nabla_\nu \phi)   
\nn
&~~~~
+  G_{5X} G^\mu{}_\nu -\frac12 G_{5XX} \big[ \delta^{\mu}_{\nu}(\Box\phi)^2 - \delta^{\mu}_{\nu}(\nd_\rho\nd_\sigma\phi)^2 
\nn
&~~~~  -2\Box\phi \nd^\mu\nd_\nu\phi +2 \nd^\mu \nd_\rho \phi \nd^\rho \nd_\nu \phi \big]   \bigg)
\nn
& ~~~~ +2G_{4X} R^{\mu}{}_{\rho} \nd^\rho \phi + G_{5X} \bigg( -\Box \phi R^\mu{}_\rho \nd^\rho\phi 
\nn
&~~~~ + R_{\rho\nu}{}^{\sigma\mu} \nd^\rho\nd_\sigma\phi \nd^\nu\phi + R_\rho{}^\sigma \nd^\rho\phi  \nd^\mu\nabla_\sigma\phi \bigg) .
\eal

We would like to check the explicit form of this current in the spherically symmetric case. To this end, we plug in a general spherically symmetric ansatz:
\bal
\ud s^2 &= -P(r) \ud t^2 +S(r) \ud r^2 + r^2 (\ud \theta^2 + \sin^2\!\theta\, \ud \varphi^2)  ,
\\
\phi &= \phi(r)  ,
\eal
which reduces the current to
\bal
\label{Jrexpr}
J^r &= 
-\frac{\phi'}{S}K_X + \frac{r(\phi')^2 P' + 4(\phi')^2R}{2rP S^2}G_{3X} 
\nn
&~~~~
+\frac{2\phi'P -2\phi'PS + 2r\phi' P'}{r^2P S^2}G_{4X} 
\nn
&~~~~ 
- \frac{2(\phi')^3P +2r(\phi')^3P' }{r^2P S^3}G_{4XX} 
\nn
&~~~~
+\frac{(\phi')^2 SP' - 3(\phi')^2 P'}{2r^2P S^3}G_{5X} 
+\frac{(\phi')^4 P' }{2r^2P S^4}G_{5XX}    ,
\eal
where a prime denotes a derivative w.r.t.~the aerial radius coordinate $r$.

It is worth emphasizing that we are assuming that $\phi$ respects the symmetries of the metric and, hence, has vanishing Lie derivatives with respect to all Killing vectors. This is the usual assumption in the literature of no-hair theorems and in particular of Ref.~\cite{Hui:2012qt}. However, as we have pointed out in Ref.~\cite{Sotiriou:2013qea}, one could choose to relax this assumption and require that only the gradient of $\phi$ has vanishing Lie derivatives. This choice can be justified by the fact that $\phi$ appears in the field equations only through its gradient, thanks to shift symmetry. Ref.~\cite{Babichev:2013cya} has discussed this option in more detail. Additionally, no-hair theorems usually concern black holes with flat asymptotics, and hairy black holes with other asymptotics in Horndeski's theory have been found \cite{Rinaldi:2012vy,Charmousis:2012dw,Anabalon:2013oea}.

\subsection{Possible evasions}

In \cite{Hui:2012qt}, it is assumed that the current should be of the form 
\be
J^r = \frac{\phi'}{S}F(\phi';g,g',g'')  ,
\ee
where $g$ represents the metric components and $F$ is an unspecified function. Comparing it to Eq.~(\ref{Jrexpr}), this seems to be the case, at least formally, for SSGG. (There actually are no terms with $g''$ in SSGG.) It is further assumed that $F$ will asymptote to a finite, nonzero constant at spatial infinity, where $S,P\to1$ and $\phi'\to0$. This is typically valid for a theory featuring a canonical kinetic term in the small field limit, which in turn is a reasonable restriction, as otherwise the scalar will be strongly coupled around the Minkowski vacuum.

We would like to explicitly examine whether these assumptions are always valid. The form of the current in Eq.~(\ref{Jrexpr}) actually leaves room for 3 different cases. Depending on the exact form of $K_X$, $G_{3X}$, $G_{4X}$, $G_{4XX}$, $G_{5X}$, $G_{5XX}$,  and performing a Laurent expansion, one has:
\begin{enumerate}
\item 
\emph{Every term in $J^r$ contains positive powers of $\phi'$.}\\
This case falls into the reign of the no-hair theorem \cite{Hui:2012qt}  as one indeed has $F\to -K_X(r\to\infty)$. 
\item 
\emph{$J^r$ contains terms with negative powers of $\phi'$.}\\
In this case the current would diverge as $\phi'\to 0$. Theories of these type, which correspond to most choices of $G_3, G_4$ and $G_5$ that are non-analytic as $X\to 0$, would not admit flat space with a trivial scalar configuration as a solution. Instead, the scalar would be always forced to have a non-trivial configuration and this would lead to violations of local Lorentz symmetry. If local Lorentz violation would be kept below experimental accuracies these theories could be interesting. However, their study and whether they lead to hairy black hole solutions goes beyond the scope of this paper (and of Ref.~\cite{Hui:2012qt}).
\item 
\emph{$J^r$ contains one or more terms with no dependence on $\phi'$, but no terms with negative powers of $\phi'$.}\\
In this case the current remains finite asymptotically and in flat space but it is not trivial to determine the asymptotic behavior of $F$. Hence, this is the case where one could find a loophole to the no-hair theorem. 
\end{enumerate}

It is rather easy to guess a choice of functions which would fall under case 3. For example, $G_5$ can be proportional to $\ln |X|$, and then the last two terms of Eq.~(\ref{Jrexpr}) will not depend on $\phi'$. However, our goal is more ambitious. We would like to identify {\em all} of the terms that fall under this category, argue conclusively that they do indeed lead to non-trivial solutions for the scalar in static, spherically symmetric spacetimes, and argue that these terms cannot be excluded from the action without loss of generality or fine-tuning.

If we require the $\phi$ equation of motion to contain a term independent of $\phi$ itself, the corresponding term in the Lagrangian should be linear in $\phi$, {\em i.e.}~of the form $\phi A[g]$ up to a total derivative, where $A[g]$ is a diffeomorphism scalar constructed from the metric and its derivatives. On the other hand, shift symmetry implies that $A$ itself should be a total derivative. Now, we want the Lagrangian term $\phi A$ in the constant $\phi$ limit to lead to a contribution to the metric equation of motion that is divergence free and with no more than second order derivatives. The only choice, by the Lovelock theorem, is then $A={\cal G}\equiv R^{\mu\nu\lambda\kappa}R_{\mu\nu\lambda\kappa}-4 R^{\mu\nu}R_{\mu\nu}+R^2$. Therefore, there is only one term one can have in the Lagrangian that would make a theory fall under case 3 above: $\phi$ has to have a linear coupling with the Gauss--Bonnet invariant\footnote{This is the case in 4D. In higher dimensions, obviously it should be the highest Lovelock invariant.}.

The last statement might seem in contradiction with the example we gave earlier, with $G_5\propto \ln |X|$. Indeed, the Gauss--Bonnet invariant does not even appear in the generalized galileon Lagrangian when given in terms of Eqs.~(\ref{ssgg})-({\ref{ssggl5}). However, this is just an issue of formalism, and it has been shown in Ref.~\cite{Kobayashi:2011nu} that the term $\alpha\, \phi\, {\cal G}$ corresponds precisely to the non-trivial choice 
\be
\label{gbg5}
K= G_3= G_4=0, \quad G_5= - 4 \alpha \ln |X|\,.
\ee
($\ln |X|$ is often written as $\ln X$ in the literature; the later case has to be understood with analytic continuation, as X can be non-positive.)

One may now check straightforwardly  that the present of this term does not allow for a trivial scalar configuration in black hole spacetimes. The current conservation equation $\nabla_\mu J^\mu=0$ becomes
\be
\Box \phi + \alpha \mc{G}=0\,,
\ee
if, for example, one makes the choices $K=X$, $G_3= G_4=0$, $G_5= - 4 \alpha \ln |X|$. $\mc{G}$ only vanishes in flat space, which directly implies that $\phi$ will have to have a non-trivial configuration in any other spacetime, including black hole spacetimes. It is worth stressing that this property will generically persist irrespective of the choices one makes for $K$,  $G_3$,  $G_4$, and $G_5$, so long as the linear coupling between $\phi$ and $\mc{G}$ is present. 

In summary, the term
\bal
\label{genevading}
\mc{L}_{\phi \rm GB} &= \alpha\phi \mathcal{G}
\nn
&= \alpha\phi (R^{\mu\nu\rho\sigma}R_{\mu\nu\rho\sigma}-4R^{\mu\nu}R_{\mu\nu} +R^2)   ,
\eal
is the only one that fall under the class of shift symmetric generalized galileons that does not lead to Lorentz invariance violations and at the same time evades the no-hair theorem of \cite{Hui:2012qt}. When this term is present black holes {\em necessarily} have hair. 

\subsection{Naturalness}
\label{sec:natu}

The choice $G_5=- 4 \alpha \ln |X|$, or equivalently the term in Eq.~(\ref{genevading}), is certainly special within the full Horndeski theory or generalized galileons. However, what we have argued above is that any theory that contains it in the action, amongst other terms, will have hairy black holes. Hence, one would need to excise this term in order to get theories in which black holes are the same as those of general relativity. Mathematically this is straightforward. But from an effective field theory's point of view this term {\em has to} be present, so long as it is not forbidden by the symmetries of the theory, as it would be generated by quantum corrections even if it were not present at the tree level. In this sense, the evasion of the no-hair theorem is quite generic. 
One can impose some internal symmetry for the scalar in order to excise  the $\phi \mathcal{G}$ term. But the choices are limited as $\phi$ is real. One possibility is to impose reflection symmetry for the scalar, $\phi\to -\phi$.
This symmetry can indeed do away with the $\phi \mathcal{G}$ term, but it also forbids the $\mc{L}_3$ and $\mc{L}_5$ terms in SSGG, thus significantly reducing the range of validity of the no-hair theorem in theory space.

\section{An explicit example: scalar-Gauss-Bonnet gravity}
\label{sec:countexam}

In the last section, we showed that the last step of the no-hair theorem of \cite{Hui:2012qt} does not go through if there is the $\phi \mathcal{G}$ term in the action. Evading the no-hair theorem is certainly essential towards having hairy black holes, but whether they truly exist is another thing. In this section, we will explicitly generate a hairy black hole solution and determine the basic nature of the scalar hair. 

We will work on the simplest model with the class of theories that have hairy black holes, that is
\be
\label{gbaction}
S=   \frac{m_P^2}{8\pi} \int\ud^4x \sqrt{-g}  \left( \frac{R}{2} - \frac12 \pd_\mu \phi \pd^\mu \phi +\alpha \phi \mathcal{G} \right) \,,
\ee
where $m_P$ is the Planck mass and $\alpha$ has dimensions of a length squared, to prove the possibility of existence. 
The Noether current associated with shift symmetry is
$J^\mu= \sqrt{-g}(\pd^\mu \phi + \alpha \tilde{\mc{G}}^\mu)$, where an expression for $\tilde{\mc{G}}^\mu$ can be found in Appendix \ref{sec:GBterm}.

The equations of motion for the metric and the scalar are respectively
\bal
\label{mEEoM}
 G_\mn &= \mathcal{T}_\mn   ,
\\ 
\label{mEEoM2}
\Box \phi +\alpha \mathcal{G} &=0   ,
\eal
where the effective energy-momentum tensor is given by
\bal
\mc{T}_\mn &= \pd_\mu\phi \pd_\nu\phi -\frac12 g_\mn (\pd\phi)^2  
\nn
&~~~~
 - \alpha (g_{\rho\mu}g_{\delta\nu}  + g_{\rho\nu}g_{\delta\mu})   \nd_\sigma\left(\pd_\gamma \phi \: \epsilon^{\gamma\delta\alpha\beta}\epsilon^{\rs\lambda\eta} R_{\lambda\eta\alpha\beta}   \right)  .
\eal 
Some more details of the derivation of the effective energy tensor are given in Appendix \ref{sec:sGBt}. It is worth pointing out that, for a covariant scalar-tensor system, the scalar equation of motion is dynamically redundant, as it is implied by the fact that the energy-momentum tensor is divergence-free, which is in turn implied by the contracted Bianchi identity. (see e.g.~\cite{Horndeski:1974wa, Frusciante:2013haa}). So it is consistent to just solve Eq.~(\ref{mEEoM}).

Before diving into the details of obtaining  black hole solutions, we note that such solutions for theories involving the Gauss-Bonnet term has been discussed before (e.g.~\cite{Wiltshire:1985us, Kanti:1995vq, Cai:2001dz, Charmousis:2008kc}). In fact, black holes with hair have been found in a theory similar to that of (\ref{gbaction}), but with $\phi \mathcal{G}$ replaced with $e^{\phi} \mathcal{G}$ \cite{Kanti:1995vq}. The models with $\phi \mathcal{G}$ and $e^{\phi} \mathcal{G}$ 
are quite distinct in the context of the no-hair proof. The former class enjoys shift symmetry for the scalar while the latter does not. 

\subsection{Spherically symmetric ansatz}
\label{sec:ssagb}

We are interested in static, spherically symmetric solutions. The most general ansatz with these symmetries in the Schwarzschild coordinates can be written as
\bal
\label{metr}
\ud s^2 &= \eta (- e^{A(r)} \ud t^2 +   e^{B(r)} \ud r^2) 
 + r^2 (\ud \theta^2 + \sin^2\theta \ud \varphi^2) ,
\\
\phi &= \phi(r)  ,
\eal
where $\eta=1$ for solutions outside the event horizon and $\eta=-1$ for  solutions inside the horizon.
Substituting this ansatz into the equations of motion (\ref{mEEoM}) and (\ref{mEEoM2}), the non-trivial components of the modified Einstein equation become
\bal 
\label{orieom1}
tt:~~~0&=16\alpha(\eta-e^{-B})\phi''+ 8 \alpha (3e^{-B}-\eta)\phi'B' 
\nn
&~~~~  +\eta r^2\phi'^2-2\eta r B' -2e^B+2\eta ,
\\
\label{orieom2}
rr:~~~0&=(e^B)^2  +12\alpha\phi'A' 
\nn
&~~~~ -\eta\left(1+rA'+4\alpha \phi'A'- \frac12 r^2\phi'^2 \right) e^B ,
\\
\label{orieom3}
\theta\theta:~~~0&=16\eta\alpha(\phi'A''+\phi''A') -2re^B A'' 
\nn
&~~~~
 +8\eta\alpha(A'-3B')A'\phi' -2re^B\phi'^2 
\nn
&~~~~ -re^B(A'-B')A' -2e^B(A'-B')   ,
\eal
and the $\phi$ equation of motion is
\bal
\label{phioeom}
\phi:~~~~0&= 8\alpha(\eta e^{-B}-1)A'' +2r^2\phi'' 
\nn
&~~~~+ r^2 (A'-B')\phi' +4\alpha(1-3\eta e^{-B})A'B'
\nn
&~~~~+4\alpha(\eta e^{-B}-1) A'^2 +4r\phi'    .
\eal
Explicit expressions for $\Box \phi$, $\mc{G}$ and the non-zero components of $G_{\mu\nu}$, $\mathcal{T}_\mn$ are given in Appendix \ref{sec:eomss}. We will mainly solve the components of the modified Einstein equation and occasionally make use of the scalar equation of motion. Algebraically solving the $rr$ component, we obtain
\be
\label{eBeom}
e^B=\frac{-\Gamma+\xi\sqrt{\Gamma^2-48\alpha\phi'A'}}{2}, \quad \xi=\pm 1  ,
\ee
where
\be
\Gamma =  -\eta\left(1+rA'+4\alpha \phi'A'- \frac12 r^2\phi'^2 \right)   .
\ee
Assuming the usual $1/r$ fall-off for the metric and $\phi$ as $r\to \infty$, as we are interested in asymptotically flat solutions, one gets $\Gamma\to -1$. This implies that we should select the $\xi=+1$ branch in Eq.~(\ref{eBeom}).\footnote{A softer fall-off would lead to the same result but would require a more convoluted argument.}

Taking a derivative of the $rr$ component with respect to $r$, one can obtain an expression for $B'(r)$. Using these expressions for $B$ and $B'$, we can re-cast the $tt$ and $\theta\theta$ components as
\bal
\label{phippeom}
\phi''(r)&= f(r,\phi',A';\alpha,\eta)   ,
\\
\label{Appeom}
A''(r)&= g(r,\phi',A';\alpha,\eta)   .
\eal
This is a closed dynamical system of ordinary differential equations (ODEs) where $\phi'(r)$ and $A'(r)$ are dynamical variables and $\alpha$ and $\eta$ are input constants. $f(\phi',A';\alpha,\eta)$ and $g(\phi',A';\alpha,\eta)$  can be straightforwardly obtained using a symbolic computation but we will refrain from displaying them here, as they are rather cumbersome.

Obtaining explicit solutions of the ODE system given by (\ref{phippeom}) and (\ref{Appeom}) analytically is a hard task. In what comes next we will follow two routes in parallel. We will seek for exact solutions numerically and we will also obtain solutions analytically in the limit where $\alpha$ is small compared to 
the characteristic length scale of the system (small coupling limit). But before doing so, we will gain some insight on the basic characteristics of the solutions we should expect to obtain by studying the approximate solutions at spatial infinity and near the horizon.

\subsection{Approximate solutions at spatial infinity}
\label{sec:appsolinfinity}

We are interested in asymptotic flat solutions, so as $r\to \infty$ we have $e^A\to 1$, $e^B\to 1$ and $\phi'\to 0$. We can then exploit shift symmetry and set $\phi|_{r\to \infty}\to 0$. In order to explore the behavior of the solution at spatial infinity, 
we expand, as a power series in $1/r$,
\bal
e^{A(r)} = 1+ \sum^{\infty}_{n=1} c^A_n/r^n\,,\\
e^{B(r)} = 1+ \sum^{\infty}_{n=1} c^B_n/r^n\,,\\
\phi(r) = \sum^{\infty}_{n=1} c^\phi_n/r^n\,,
\eal
where $c^A_n$, $c^B_n$ and $c^\phi_n$ are constants to be determined. Substituting these series into Eqs.~(\ref{orieom1}-\ref{orieom3}), we can perturbatively solve to get, to order $1/r^4$,
\bal
\label{eAinf}
e^A(r)&=1-\frac{2M}{r}+\frac{M P^2}{6r^3}
\nn
&~~~~
+\frac{M^2 P^2+24\alpha M P}{3r^4}+\mc{O}\left(\frac{1}{r^5}\right)   ,
\\
\label{eBinf}
e^{B(r)}&=1+\frac{2M}{r}+\frac{8M^2 -P^2}{2r^2}+\frac{16 M^3 - 5 M P^2}{2r^3}
\nn
&~~~~+\frac{192{M}^{4}-104{M}^{2}{P}^{2}-192\alpha MP+3{P}^{4}}{12r^4} 
\nn
&~~~~
+\mc{O}\left(\frac{1}{r^5}\right)   ,
\\
\label{phiinf}
\phi(r)&=\frac{P}{r}+\frac{MP}{r^2}+\frac{16M^2 P-P^3}{12r^3}
\nn
&~~~~
+\frac{6M^3P-12\alpha M^2-MP^3}{3r^4}+\mc{O}\left(\frac{1}{r^5}\right)  ,
\eal
where we have used geometrized units with $m_P$ set to 1. As can be defined by usual boundary integrals, here $M$ is the ADM mass of the potentially existing black hole, and $P$ is the scalar charge, which potentially is the extra scalar ``hair''. Hence, generically there will be a 2-parameter family of solutions. This could have been expected since, thanks to spherical symmetry, Eqs.~(\ref{phippeom}) and (\ref{Appeom})  can be thought of as an initial value problem with $r$ playing the role of ``time". On a given $r$, one needs to provide 2 pieces of initial data, the values of $A'$ and $\phi'$ there.

 \subsection{Approximate solutions near the horizon}
 \label{sec:appsolhorizon}

We are interested in black hole solutions, so we will assume that there is an event horizon at $r=r_h$.  Thus, we should have $e^A|_{r\to r_h^+}\to 0^+$ outside the horizon ($e^A|_{r\to r_h^-}\to 0^-$ inside the horizon), where $r\to r_h^+$ ($r\to r_h^-$) means approaching the horizon from outside (inside) the horizon.  That is, we have $A'|_{r\to r_h^+}\to +\infty$ when $\eta=1$ ($A'|_{r\to r_h^-}\to -\infty$ when $\eta=-1$). 

As $r\to r_h^{\pm}$ and $A'|_{r\to r_h^{\pm}} \to \pm \infty$, we can see from Eq.~(\ref{eBeom}) that 
\be
e^B \to \frac{1+ {\rm sign}(r+4 \alpha \phi'_h)}{2} (r +4 \alpha \phi'_h)\eta A' |_{r\to r_h^{\pm}}\,,
\ee
where we have assumed that $\phi'$ is continuous near the event horizon and thus the limit $\phi'_h\equiv \phi'|_{r\to r_h^{\pm}}$ exists. $e^B$ should diverge near the horizon, which implies that 
\be
r_h  + 4\alpha \phi'_h >0\,.
\ee

Then Eq.~(\ref{eBeom}) can be expanded near the horizon as
\bal
e^B&=\eta \left[(4\alpha \phi'+r) A'  -\left(\frac{r^2\phi'^2}{2} + \frac{8\alpha \phi'-r}{4\alpha \phi'+r}\right) \right]
\nn
&~~~~
+\mc{O}\left( \frac{1}{A'}\right)   .
\eal
Substituting this expression into Eq.~(\ref{phippeom}) and (\ref{Appeom}), we have
\bal
\label{phippeomexp}
\phi''&=-\frac{(4\alpha \phi'+r)[r^2\phi'(4\alpha \phi'+r) + 12\alpha]}{r^3(4\alpha \phi'+r) - 96\alpha^2} A' 
\nn
&~~~~ +\mc{O}(1)    ,
\\
\label{Appeomexp}
A''&= -\frac{r^2(4\alpha \phi'+r)^2-48\alpha^2}{r^3(4\alpha \phi'+r) - 96\alpha^2}A'^2+\mc{O}(A')  .
\eal

For $\phi''$ to be finite on the horizon one need to have $r_h^2\phi'_h(4\alpha \phi'_h+r_h) + 12\alpha=0$, which yields
\be
\label{phiph}
\phi'_h = \frac{-r_h^2 \pm \sqrt{r_h^4 - 192 \alpha^2}}{8\alpha r_h}.
\ee
This equation can be interpreted as a regularity condition for $\phi$ on the horizon. As discussed previously, one can formulate the problem of generating a solution to Eqs.~(\ref{phippeom}) and (\ref{Appeom}) as an initial value problem, with $r$ playing the role of ``time''. Then the values of $\phi'$ and $A'$ on the initial hypersurface would serve as initial data. For a black hole solution it would be most convenient to choose the horizon as the initial hypersurface, as this guaranties that a horizon actually exists. However, Eq.~(\ref{phiph}) fixes the value of $\phi'$ there. At they same time $A'$ diverges on the horizon by definition, so the only piece of data that is free to choose is $r_h$, the actual location of the horizon (which will then determine the ADM mass of the black hole). Hence, one expects that there will be a 1-parameter family of black hole solutions. This is to be contrasted with the asymptotic behavior of a generic solutions, explored in the previous section, which was found to have 2 independent charges. We can conclude that regularity on the horizon fixes the scalar charge in terms of the mass of the black hole. 

For $\phi'_h$ to be real, one additionally needs that
\be
\label{conalpha}
r_h >r_h^M\equiv  2\sqrt{2\sqrt{3}|\alpha|}   .
\ee
As we shall see in more detail below, $r_h^M\equiv 2\sqrt{2\sqrt{3}|\alpha|}$ is the minimum size a black hole can have for a given $\alpha$. Solutions that do not meet this criterion have naked singularities.
Note that this condition is consistent with $4\alpha\phi'_h+r_h > 0$, as it should. 

\subsection{Numerical implementation and a fiducial solution}
\label{sec:nfs}

We have already mentioned that the strategy we intend to follow in order to numerically generate a solution will be to start from the horizon and integrate outwards and inwards. For a given value of the coupling $\alpha$, different values for the horizon radius would correspond to solutions with a different mass (as we will demonstrate in detail later). One could decide to fix the dimensionless parameter $\alpha/r_h^2$ and generate the solution for $r_h=1$. Then, solutions with different horizon radii (and masses) could be simply generated by appropriately rescaling the radial coordinate. It should be stressed that this process generates (after the rescaling) a family of solutions with, not only different masses, but also different couplings $\alpha$ (as it is $\alpha/r_h^2$ that is held fixed). However, it allows one to explore a range of couplings and masses by scanning  only the 1-dimensional parameter space of $\alpha/r_h^2$, instead of the 2-dimensional parameter space of $(r_h, \alpha)$. 

A numerical issue that one has to face is that the regularity constraint of Eq.~(\ref{phiph}) cannot be imposed numerically with any sensible accuracy. This should be clear from the fact that the first term on the r.h.s. of Eq.~(\ref{phippeomexp}) is seen as an unresolvable $0\times \infty$ by any finite-accuracy calculator. To circumvent this problem, one can generate a perturbative solution  around $r=r_h$. Then one can make use of this perturbative solutions in order to calculate $\phi'$ and $A'$ at a distance $\epsilon r_h$ from the horizon, both inwards and outwards, and start the numeric integration from these radii, $r_\pm$.  This is the method  we have followed. We have actually set $\epsilon=10^{-9}$ and used linear perturbation theory for calculating $A'(r_\pm)$, whereas we have taken for $\phi'(r_\pm)$ to be simply the positive-sign root of Eq.~(\ref{phiph}). Choosing the negative-sign root would in principle correspond to a second branch of solutions, but in practice numerics do not generate a sensible solution. Our approximation for $A'(r_\pm)$ and $\phi'(r_\pm)$ might seem to be crude. However, one can use the continuity of $\phi''$ as one approaches the horizon from the two sides as an indicator of accuracy for our approximation, and we find the values to match to order $10^{-9}$.  We use the Maple ODE solver \emph{rk45} with the error control parameters $abserr$ and $relerr$ set to $10^{-12}$ and $10^{-10}$ respectively and the environment viable $Digits$ set to $15$.

It is worth noting that action (\ref{gbaction}) is invariant under the combined operation $\phi\to -\phi$ and $\alpha\to-\alpha$. Therefore, we can choose $\alpha >0$ without lost of generality, as the solutions with $\alpha<0$ can be simply obtained from the corresponding $\alpha>0$ solutions with $\phi$ replaced by  $-\phi$.

\bwt
\begin{center}
\begin{figure}[ht]
\centering
\subfigure[~$\phi(r)$]{%
\includegraphics[width=0.3\linewidth]{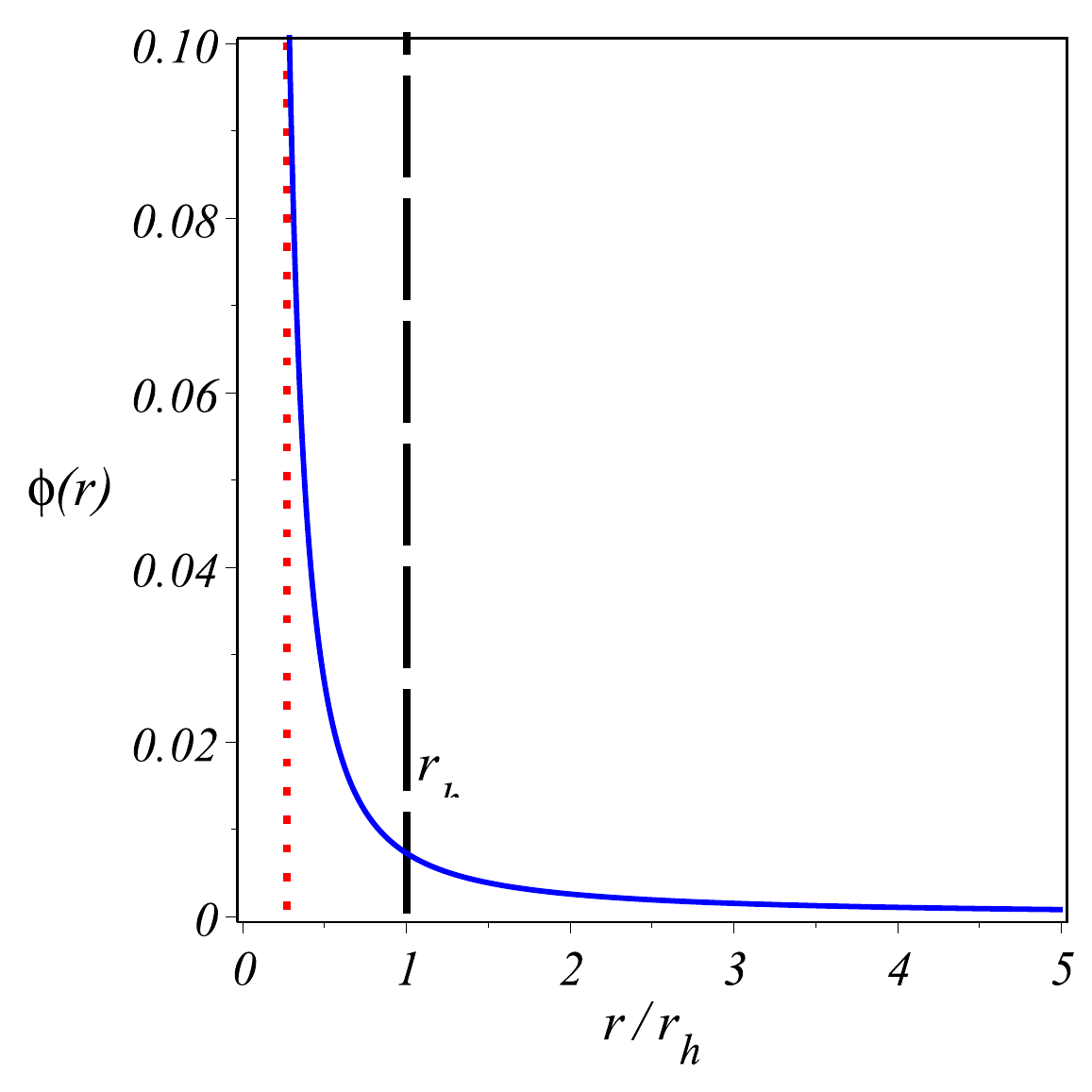}
\label{fig:phiofr}}
\subfigure[~$A(r)$]{%
\includegraphics[width=0.3\linewidth]{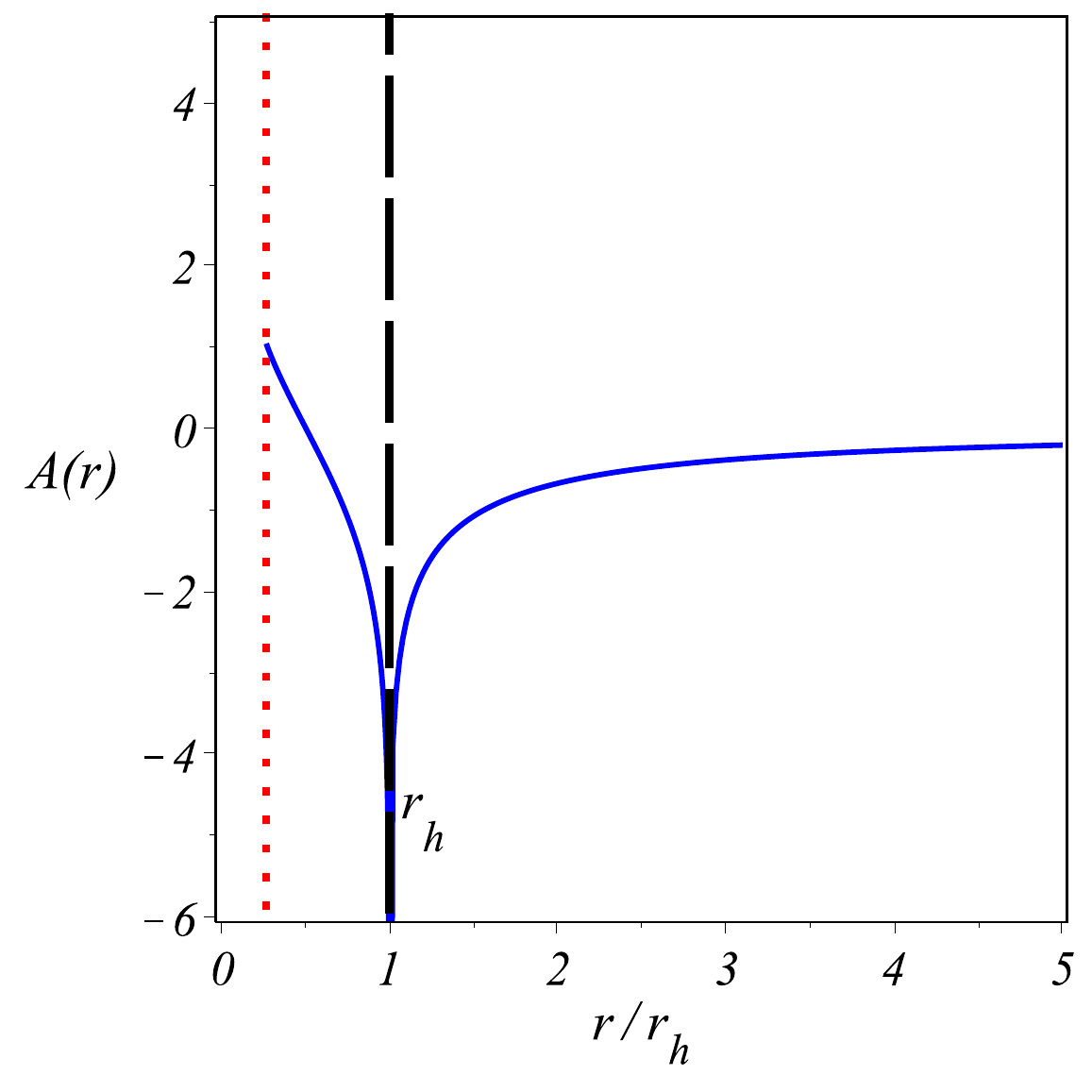}
\label{fig:Aofr}}
\subfigure[~$\ln (\mc{G}r_h^4)$]{%
\includegraphics[width=0.3\linewidth]{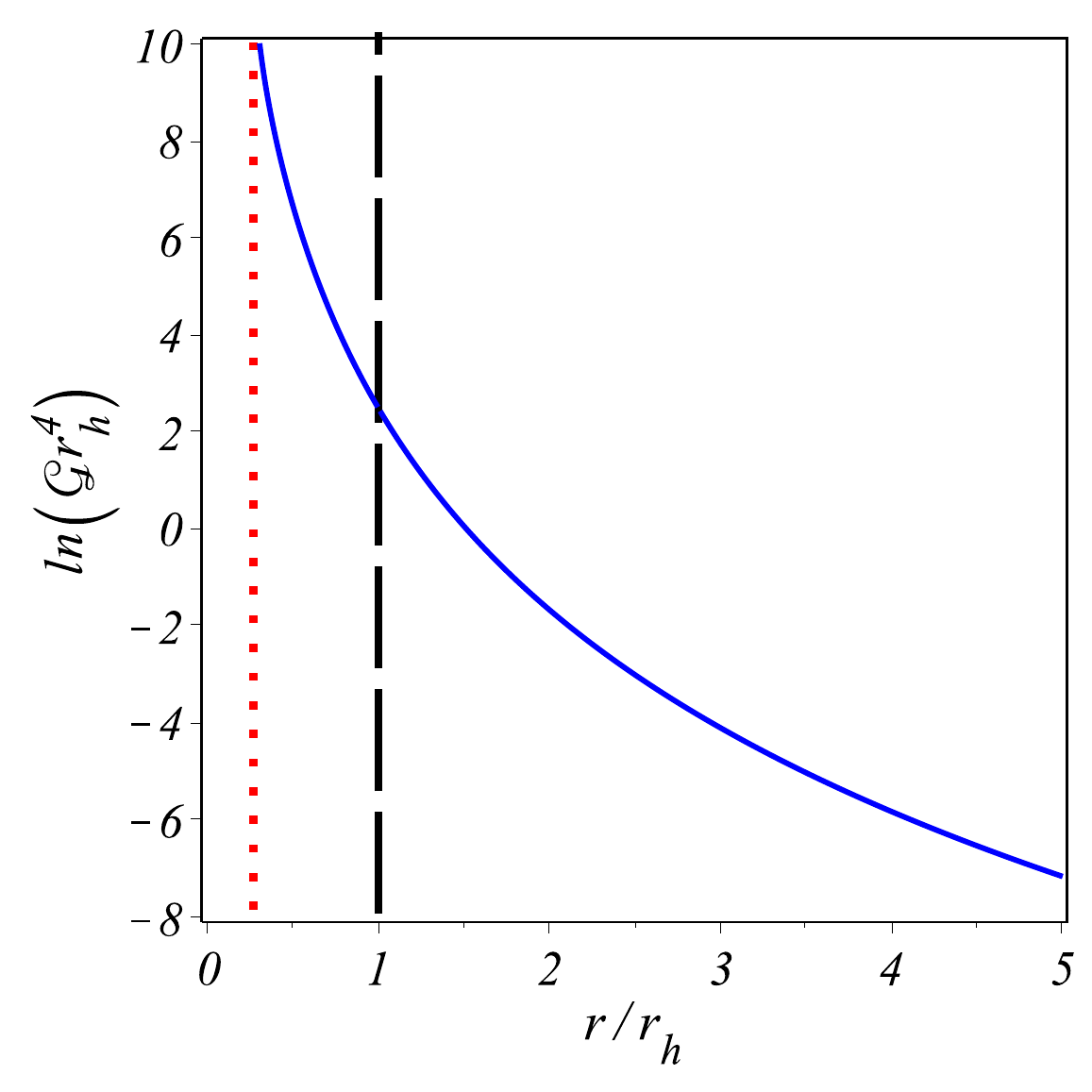}
\label{fig:lnG}}
\caption{A fiducial solution with $\alpha/r_h^2=0.001$. $r_h$ is the horizon radius (black, long-dashed vertical line), which is used as the unit of length. There is a finite radius singularity ($r\approx 0.2689 r_h$), indicated by the red, dotted vertical line.  (a):  The scalar field $\phi(r)$, normalized to 0 at spatial infinity; (b): The metric component $g_{00}=-\eta e^{A(r)}$ ($\eta=1$ for $r>r_h$ and $\eta=-1$ for $r<r_h$); (c):  The Gauss-Bonnet invariant $\mc{G}$.}
\label{fig:fidsol}
\end{figure}
\end{center}
\ewt

As a preview to our results, we show a few characteristic quantities of a fiducial black hole solution with $\alpha /r_h^2=0.001$ in Fig.~\ref{fig:fidsol}. The most prominent feature is the presence of a finite surface singularity, which is nevertheless cloaked by the horizon if $r_h >r_h^M\equiv  2\sqrt{2\sqrt{3}|\alpha|}$. Finite surface singularities are not surprising for a theory involving the Gauss-Bonnet term. For example, they have also been observed in Gauss-Bonnet gravity in higher dimensions without coupling to a scalar \cite{Wiltshire:1985us, Cai:2001dz} or Gauss-Bonnet gravity coupling to non-shift-symmetric scalar \cite{Kanti:1995vq}. We will return to the issue of the finite area singularity in the next Section, after deriving first perturbative solutions in the small coupling limit. This will give us the opportunity to compare  these perturbative solutions with the full numerical one and understand the role of non-perturbative effects.

\subsection{Perturbative solutions in the small coupling limit}
\label{sec:hasi}

We now turn out attention to finding a perturbative solution in the small coupling limit. 
We can consider the horizon radius $r_h$ as  a characteristic length associated with our solution. Then one can define the dimensionless parameter $\tilde{\alpha}\equiv \alpha/r_h^2$ and attempt to generate a solution perturbative in the limit $\tilde{\alpha}\ll 1$.

If $\tilde\alpha$ were zero, the theory would reduce to General Relativity minimally coupled to a scalar and we would have the standard Schwarzschild solution. Adding the scalar-Gauss-Bonnet term with a small $\tilde\alpha$ induces a perturbatively deformed Schwarzschild solution, which we assume to be
\bal
\label{pertmetric}
\ud s^2 &= - \left(1-\frac{2m}{r}\right) \left(1+\sum_{n=1}A_n(r)\tilde{\alpha}^n\right)^2 \ud t^2 
\nn
&~~~~
+ \left(1-\frac{2m}{r}\right)^{-1} \left(1+\sum_{n=1}B_n (r)\tilde{\alpha}^n\right)^2 \ud r^2
\nn
&~~~~
 +r^2 (\ud \theta^2+\sin^2\theta\ud \varphi^2)   .
\eal
The scalar is assumed to be
\be
\label{pertphi}
\phi'=\sum_{n=1}\phi'_n(r) \tilde{\alpha}^n    .
\ee
Note that we used $m$ in the metric, rather than $M$ (the mass of the black hole as measured by a far away observable), because they may be different. 

Substituting these ansatz into the equations of motion Eqs.~(\ref{orieom1}), (\ref{orieom2}), (\ref{orieom3}) and (\ref{phioeom}), and solving order by order in $\tilde{\alpha}$ one obtains a perturbative solution. In a slight abuse of notation, in what follows we will set $r_h=1$ and drop the twiddle, identifying $\alpha$ and $\tilde{\alpha}$. $r$ and $m$ should be then effectively measured in units of $r_h$, similarly to our numerical solutions.

Note that the approach we follow in the section has been followed in Ref.~\cite{Yunes:2011we}  for a more general theory that includes general couplings between the scalar and the Chern--Simons and Gauss--Bonnet invariants. However, a weak field limit approximation was additionally employed and it was assumed that the coupling functions are dominated by the linear term in this limit. Under these assumptions the Chern-Simons term gives no contribution to second order in the coupling and, hence, the solutions we will present below are in full agreement with those of Ref.~\cite{Yunes:2011we}.\footnote{In \cite{Yunes:2011we}, a consistent condition for the perturbative regime to be valid has also been established, which translated to $\alpha^2/r_h^4<15/1444$ in our formalism. Our consistency condition based on the full solution, i.e., Eq.~(\ref{conalpha}), is stronger.}

\subsubsection{1st order}

We had stated earlier that one could obtain a consistent solution by solving the modified Einstein equation alone and ignoring the scalar equation. However, in the perturbative treatment it is significantly simpler to solve the scalar equation itself in order to determine $\phi$ at each order. 
The first order equations of motion are
\bal
(r-2m)B'_1+B_1 &=0     ,
\\
(r-2m)A'_1-B_1 &=0      ,
\\
r(r-2m) A''_1+(r+m)A'_1-(r-m)B'_1 &=0     ,
\\
\label{forderscalar}
(r-2m)r^5\phi''_1 + 2(r-m)r^4\phi'_1+48m^2 &=0     ,
\eal
where Eq.~(\ref{forderscalar}) is the scalar equation of motion. Imposing the boundary conditions that $A_1$, $B_1$ and $\phi'_1$ vanish at spatial infinity, these equations can be easily solved:
\bal
A_1 &= \frac{c_1}{r-2m}     ,
\\
B_1 &= -\frac{c_1}{r-2m}     ,
\\
\phi'_1 &= {\frac {16{m}^{2}-c_2{r}^{3}}{{r}^{4} \left( r-2m \right) }}     ,
\eal
where $c_1$ and $c_2$ are undetermined constants. We also have boundary conditions at the horizon and the requirement that the perturbative expansion should remain under control, which leads to 
\bal
c_1 =0 ,~~~~ c_2 =\frac{2}{m} ,
\eal
and so
\be
A_1=0,~~~~B_1=0,~~~~\phi'_1= -\frac{2(r^2+2mr+4m^2)}{mr^4}.
\ee
Therefore, to leading perturbative order, the metric is not modified by the scalar configuration. This is the solution that is given in Ref.~\cite{Sotiriou:2013qea}. By comparing with the asymptotic solution at $r\to +\infty$, the ADM mass and the scalar charge at this order are
\be
M=m, ~~~~ P=  \frac{2\alpha}{M}  .
\ee
As expected, the scalar charge is fixed for a given mass, due to the regularity condition at the horizon. It is remarkable that, as the mass increases, the scalar charge has to decrease for a given coupling $\alpha$.

Before going further, one should mention that, depending on how action (\ref{gbaction}) arises from a more fundamental theory, it might or it might not make sense to go further in perturbation theory. In particular, suppose $\alpha$ is  some order parameter and the action is the product of a truncation to order $\alpha$. Then clearly one cannot trust solutions beyond that order, as terms of order $\alpha^2$ have been neglected in the action (and field equations as a consequence). In this case, the numerical solution discussed in the previous section is not particularly useful, as it would anyway be valid only in the range in which it agrees with the perturbative solution just presented. On the other hand, if 
the action is taken to be exact and not the product of a truncation, then it is worth going to next-to-leading order in perturbation theory in order to obtain a correction to the metric and also be able to compare with the numerical solution (which would then be the full solution).

\subsubsection{2nd order}

Proceeding to the next order with the help of the first order solutions, the $tt$ and $rr$ components of the modified Einstein equation and the scalar equation of motion can be cast as, respectively, 
\bal
4r^7(r-2m)^2 [(r-2m)B'_2 +B_2]
-\frac{4}{m^2} {r}^{7} &
\nn
+\frac{8}{m}{r}^{6}-192{r}^{5} +768{m}{r}^{4}-768{m}^{2}{
r}^{3} &
\nn
+3072{m}^{3}{r}^{2}-12032{m}^{4}r+11776{m}^{5}&=0     ,
\\
4r^7(r-2m)^2 [(2m-r)A'_2 +B_2]  + \frac{4}{m^2}{r}^{7} &
\nn
-\frac8m{r}^{6}- 64{r}^{5}+256{m}{r}^{4}-256{m}^{2}{r}^
{3}  &
\nn
+512{m}^{3}{r}^{2}-2304{m}^{4}r+2560{m}^{5} &=0    ,
\\
 r(r-2m)\phi'_2 +2(r-m)\phi_2&=0   .
\eal
The equations can be solved to give 
\bal
A_2&= \frac{r^2}{15m(r-2m)^2}  \bigg( 15{\frac {{c_3}m}{r}}-30{\frac {{c_3
}{m}^{2}}{{r}^{2}}}+\frac{5}{r^3}+120{\frac {m}{{r}^{4}}}
\nn
&~~~~  -194{
\frac {{m}^{2}}{{r}^{5}}}-36{\frac {{m}^{3}}{{r}^{6}}}-592{\frac 
{{m}^{4}}{{r}^{7}}}+800{\frac {{m}^{5}}{{r}^{8}}}\bigg)     ,
\\
B_2&=\frac{-1}{r-2m}\bigg({c_3}+{\frac {1}{{m}^{2}r}}+{\frac {1}{m{r}^{2}}}+{\frac {52}{3r^3}}
+2{\frac {m}{{r}^{4}}}
\nn
&~~~~ +{\frac {16}{5}}{\frac {{m}^{2}}{{r}^{5}}}-{
\frac {368}{3}}{\frac {{m}^{3}}{{r}^{6}}}\bigg)     ,
\\
\phi_2 &= \frac{c_4}{r(r-2m)},
\eal
where $c_3$ and $c_4$ are integration constants. Now, we require $A_2$, $B_2$ and $\phi_2$ to be controlled perturbations at $r=2m$, this imposes
\be
c_3 =-\frac{49}{40m^3}   ,~~~~c_4 =0   ,
\ee
and so
\bal
\label{A2exp}
A_2&= -\frac{49}{40m^3r} -\frac{49}{20m^2r^2} - \frac{137}{30mr^3} -\frac{7}{15r^4}
\nn
&~~~~ +\frac{52m}{15r^5}+\frac{40m^2}{3r^6}  ,
\\
\label{B2exp}
B_2&= \frac{49}{40m^3r} +\frac{29}{20m^2r^2} + \frac{19}{10mr^3}  - \frac{203}{15r^4}
\nn
&~~~~  -\frac{436m}{15r^5}-\frac{184m^2}{3r^6} , 
\\
\phi_2 &=0   .
\eal
By comparing to the asymptotic solution as $r\to +\infty$, the ADM mass and the scalar charge up to 2nd order are
\be
\label{Mcorrected}
M=m \left(1+\frac{49\alpha^2}{40m^4}\right), ~~~~ P=\frac{2\alpha}{M}   .
\ee

Now, the Ricci scalar and the Gauss-Bonnet invariant can be computed analytically up to $\mc{O}(\alpha^2)$:
 \bal
 \label{R2exp}
 R &=
  \bigg( \frac{16}{{r}^6}-{\frac {32m}{{r}^{7}}}-{\frac {64{m}^{2}}{{r
}^{8}}}+{\frac {4}{{r}^{4}{m}^{2}}}+{\frac {8}{{r}^{5}m}} -{\frac {128{m}^{3}}{{r}^{9}}} \bigg) {\alpha}^{2}
\nn
&~~~~  +\mc{O} \left( {\alpha}^{3} \right) 
 \\
 \label{GB2exp}
 \mc{G} &=  
 {\frac {48{m}^{2}}{{r}^{6}}}+ \bigg( {\frac {588}{5{m}^{2}{r}^{6}}} -{\frac {64}{m{r}^{7}}}-\frac{32}
{r^8}-{\frac {4608m}{{r}^{9}}}-{\frac {448{m}^{2}}{{r}^{10}}}
\nn
&~~~~ -{\frac {4096{m}^{3}}{5{r}^{11}}}+{\frac {53760{m}^{4}
}{{r}^{12}}}\bigg) {\alpha}
^{2}+\mc{O} \left( {\alpha}^{3} \right) 
 \eal
From Eqs.~(\ref{R2exp}) and (\ref{GB2exp}), we see that the feature of the finite radius singularity is not captured in the perturbative solution up to $\mc{O}(\alpha^2)$, which is the first correction to the Schwarzschild geometry. See Fig.~\ref{fig:Gcompare}, for example, for a comparison of the Gauss-Bonnet invariant between the numerical full and perturbative solutions for a fiducial $\alpha/r_h^2$. 

However, since $A_2$, $B_2$ and $\mc{G}^{(2)}$ (the $\alpha^2$ order of $\mc{G}$) tend to infinity as $r\to 0$,  the perturbative expansion in $\alpha$ breaks down at some small enough $r$ where either $\alpha^2A_2$ or $\alpha^2B_2$ becomes $\mc{O}(1)$, or  $\mc{G}^{(2)}$ becomes $\mc{O}({{48{m}^{2}}/{{r}^{6}}})$. It turns out that $\mc{G}^{(2)}$ becomes $\mc{O}({{48{m}^{2}}/{{r}^{6}}})$ first when $r$ runs to small values for a black hole solution. In Fig.~(\ref{fig:finiteSingularity}), we have also plotted the radius for which $|\mc{G}^{(2)}|={{48{m}^{2}}/{{r}^{6}}}$. Presumably, perturbation theory actually breaks down before $|\mc{G}^{(2)}|$ reaches ${{48{m}^{2}}/{{r}^{6}}}$, but we expect the radius at which the two quantities are equal to closely track the finite area singularity radius. 

\begin{figure}
\includegraphics[height=2.7in,width=2.7in]{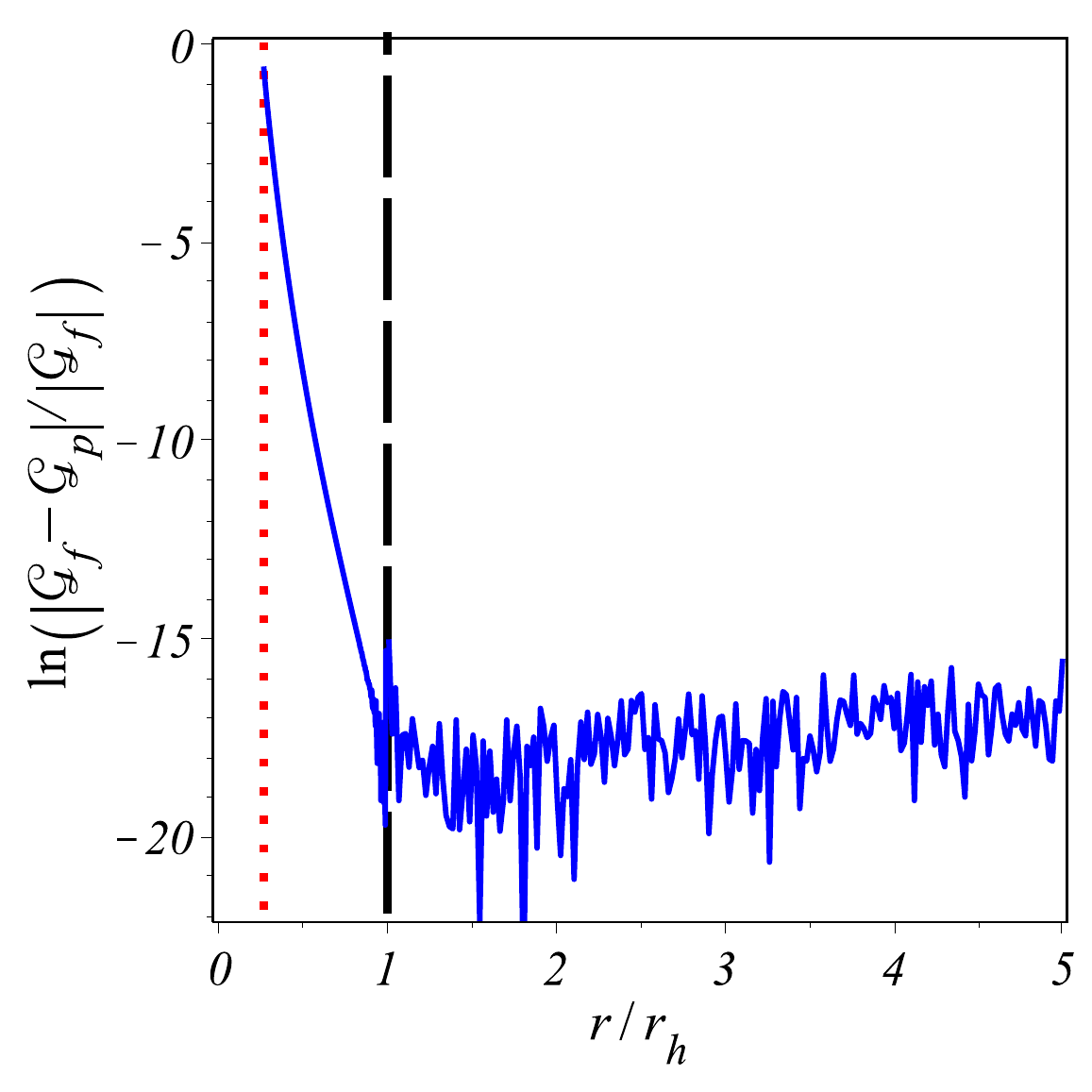}
\caption{Comparison of the Gauss-Bonnet invariant between the numerical full solution and the $\mc{O}(\alpha^2)$ perturbative solution. $\alpha$ is chosen as $\alpha/r_h^2=0.001$, where $r_h$ is the horizon radius (black, long-dashed vertical line). The red, dotted vertical line ($r\approx 0.2689r_h$) is the finite radius singularity of the full solution.} 
\label{fig:Gcompare}
\end{figure}

\begin{figure}
\includegraphics[height=2.7in,width=2.7in]{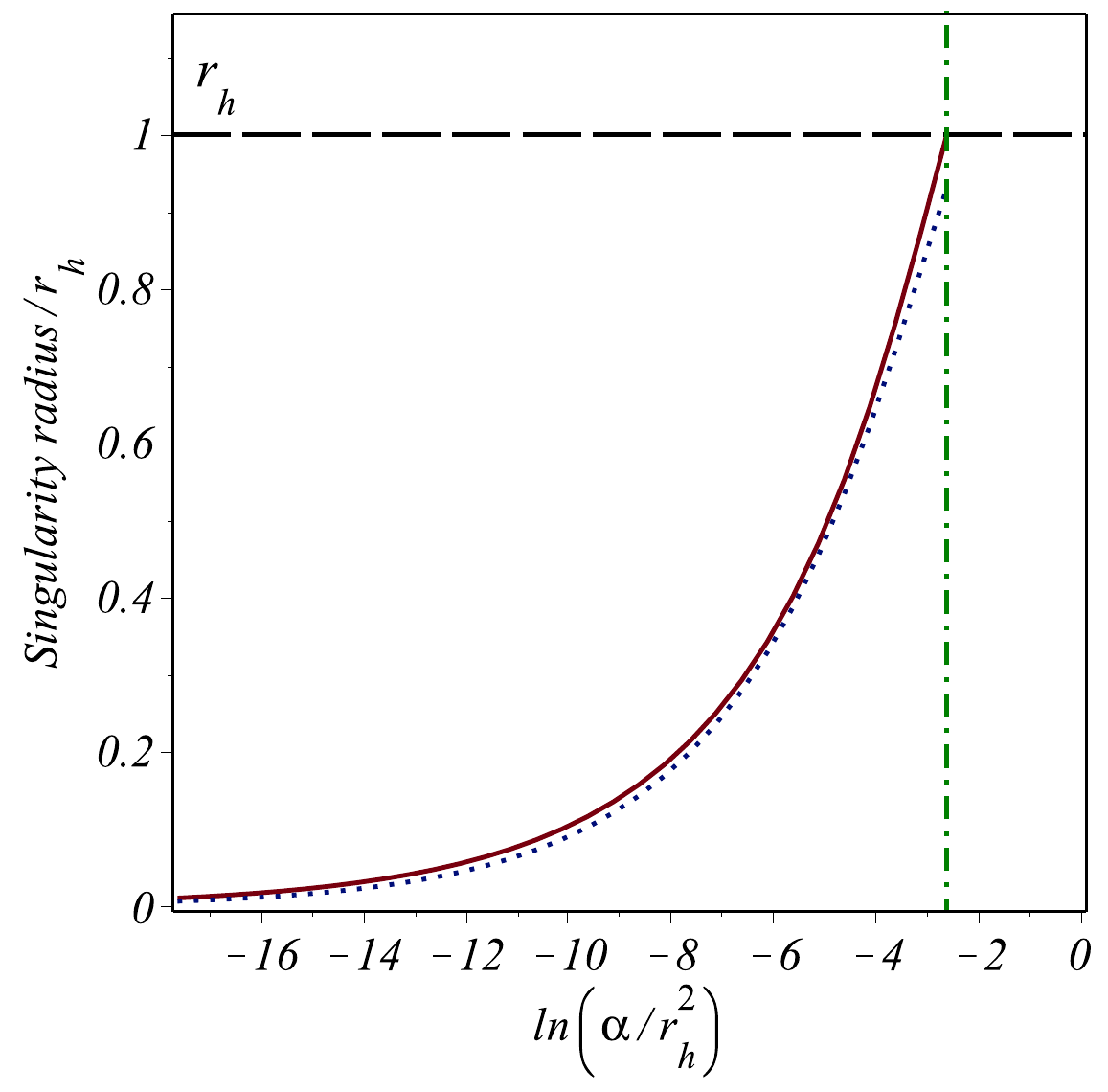}
\caption{The red, solid line is the radius of the finite singularity surface as a (numerical) function of the Gauss-Bonnet coupling $\alpha$, computed from the full numerical solution. $r_h$ is the horizon radius (black, long-dashed horizontal line). The green, dot-dashed vertical line is the  largest value $\alpha/r_h^2$ can take without the solution having a naked singularity. The blue, dotted line is the radius where the perturbation solution in Section \ref{sec:hasi} breaks down, i.e., where the $\mc{O}(\alpha^2)$ correction in Eq.~(\ref{GB2exp}) becomes comparable with the leading order.} \label{fig:finiteSingularity}
\end{figure}

\subsection{Phenomenology}
\label{sec:phenome}

Astronomical observations of black holes and their vicinities have been getting more and more precise and are expected to become an important way to detect possible deviations from GR in the near future. In this section we will compute some common observational quantities for the black hole solutions we have generated and we will compare them to those of the Schwarzschild black hole. These deviations can be used to constrain  the linear scalar Gauss-Bonnet coupling $\phi\mc{G}$ through confrontation with  current and near future experiments. Such constraints can actually be thought of as constraining the possibility of having black holes with scalar hair in SSGG, as we have shown that the linear scalar Gauss-Bonnet coupling is essential for the existence of scalar hair.

Following the lines of Ref.~\cite{Barausse:2011pu}, we will compute, for different values of $\alpha$, the innermost stable circular orbit (ISCO) angular frequency, the maximum redshift from the ISCO and the impact parameter of the circular photon orbit. We will provide definitions for these quantities shortly (see e.g.~\cite{Hartle:2003yu} for an discussion on these observables).
We will also compute the gravitational radius, defined as
\be
r_g = 2M
\ee 
where $M$ is the ADM mass of the black hole, extracted at spatial infinity as $g_{00}=1-r_g/r+\mc{O}(1/r^2)$. For the Schwarzschild solution, this is the same as $r_h$. But $r_h$ and $r_g$ do not generically coincide for black holes that are not solutions of GR.

For a massive test particle, there exists the innermost stable circular orbit around a spherical black hole. This orbit occurs when the maximum and the minimum of the effective potential for the radial motion become degenerate. The ISCO radius $r_{ISCO}$ can be determined by the following equation, using the metric (\ref{metr}) with $\eta=1$, 
\be
\label{iscorad}
3A'-rA'^2+rA'' = 0 .
\ee
The ISCO angular frequency is given by
\be
\omega_{ISCO} = \left. \sqrt{\frac{e^{A}A'}{2r}} \right|_{ISCO}  .
\ee   
A photon emitted by a source at the ISCO is redshifted when observed at infinity. The maximum redshift for a photo emitted from the ISCO is given by
\be
z_{M} = \left. e^{-\frac{A}{2}} \frac{\sqrt{2}+\sqrt{A'r}}{\sqrt{2-A' r}} \right|_{ISCO} -1 .
\ee
A photon can also have a circular orbit, which occurs at
\be
\label{phrad}
-2+rA'=0   .
\ee
The impact parameter for the photon circular orbit is given by 
\be
b_{ph} = \left. \frac{r}{\sqrt{e^A}} \right|_{ph} ,
\ee
and its angular frequency is simply $\omega_{ph}=1/b_{ph}$. 

In Figs.~(\ref{fig:rg}) and (\ref{fig:rISCO}) we show the fractional deviation of the gravitational radius $r_g$ and the ISCO radius from the horizon radius $r_h$ {and the GR ISCO radius $r=3r_h$ respectively}. Figs.~(\ref{fig:wISCOrg}), (\ref{fig:zM}) and (\ref{fig:bphrg}) are plots of the fractional deviations of  a certain quantity from the value it would have for a Schwarzschild black hole for different values of the coupling $\alpha$. Fig.~(\ref{fig:wISCOrg}) is for the dimensionless quantity $\omega_{ISCO}r_g$, Fig.~(\ref{fig:zM}) for  
$z_M$ and Fig.~(\ref{fig:bphrg}) for the dimensionless quantity $b_{ph}/r_g$. From these plots, one can clearly see that for the spherical black hole solution the deviations from GR induced by the scalar Gauss-Bonnet coupling is negligibly small, as long as we impose that the finite radius singularity be cloaked by the horizon. For the extreme case where the finite radius singularity is barely cloaked by the horizon, the deviations from GR for $\omega_{ISCO}r_g$, $z_M$ and $b_{ph}/r_g$ are less than $\sim 5\%$.

\begin{figure}
\includegraphics[height=2.7in,width=2.7in]{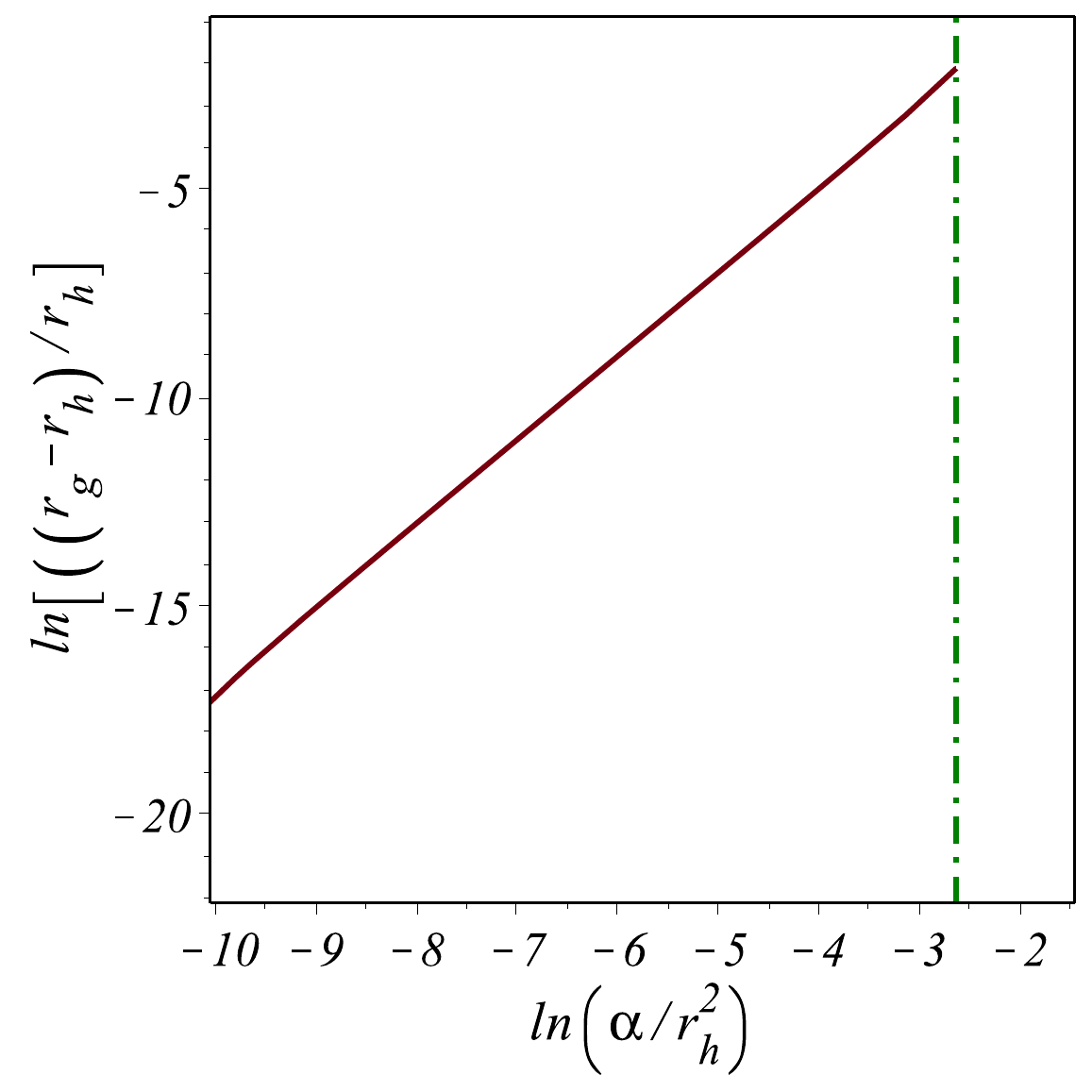}
\caption{Comparison between the gravitational radius $r_g=2M$, where $M$ is the ADM mass of the black hole  and the horizon radius $r_h$. The green, dot-dashed vertical line is the largest value  $\alpha/r_h^2$  can take without the solution having a naked singularity.} 
\label{fig:rg}
\end{figure}

\begin{figure}
\includegraphics[height=2.7in,width=2.7in]{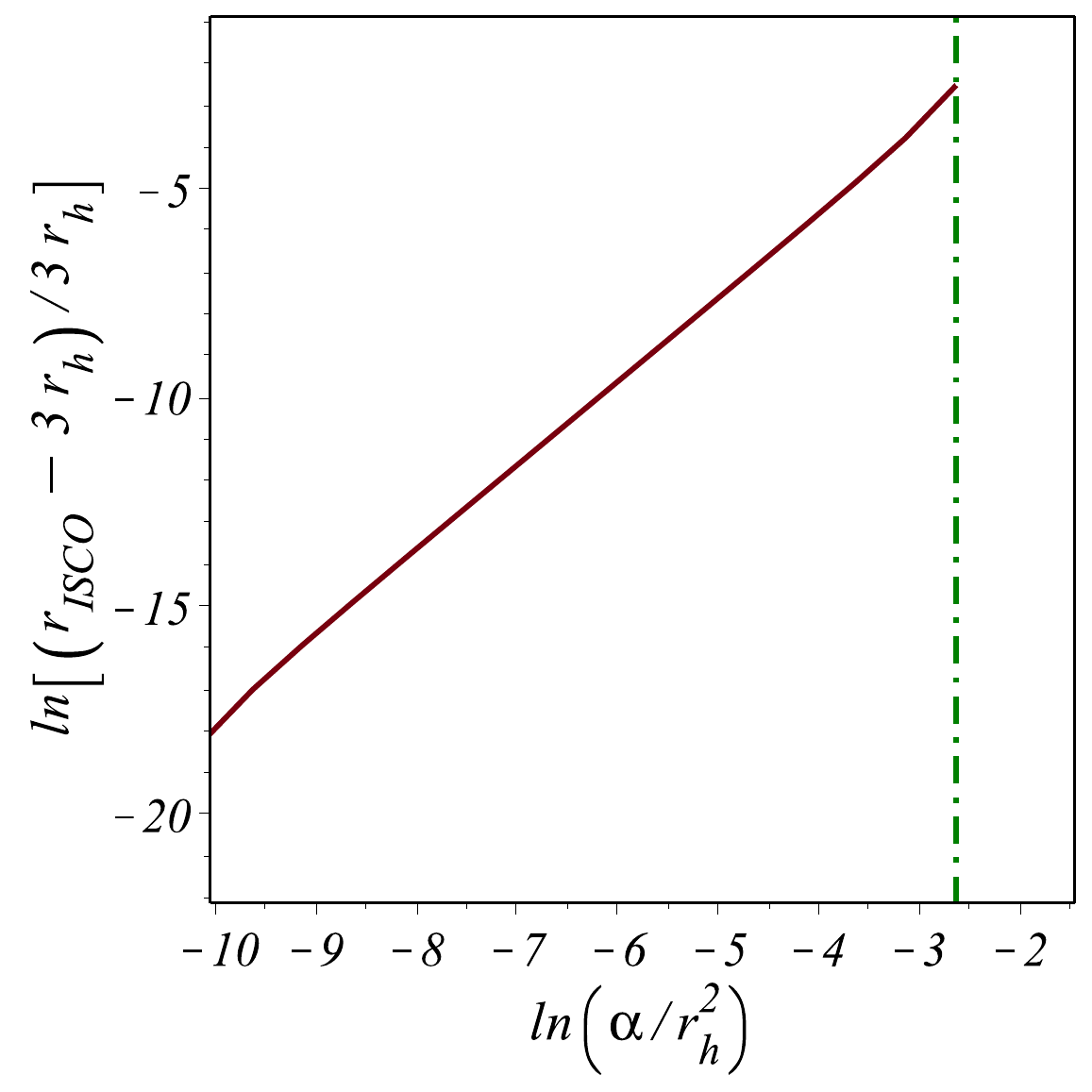}
\caption{Comparison between the ISCO radius $r_{ISCO}$ and the GR ISCO radius $3r_h$. The green, dot-dashed vertical line is the  largest value  $\alpha/r_h^2$  can take without the solution having a naked singularity.} 
\label{fig:rISCO}
\end{figure}

\begin{figure}
\includegraphics[height=2.7in,width=2.7in]{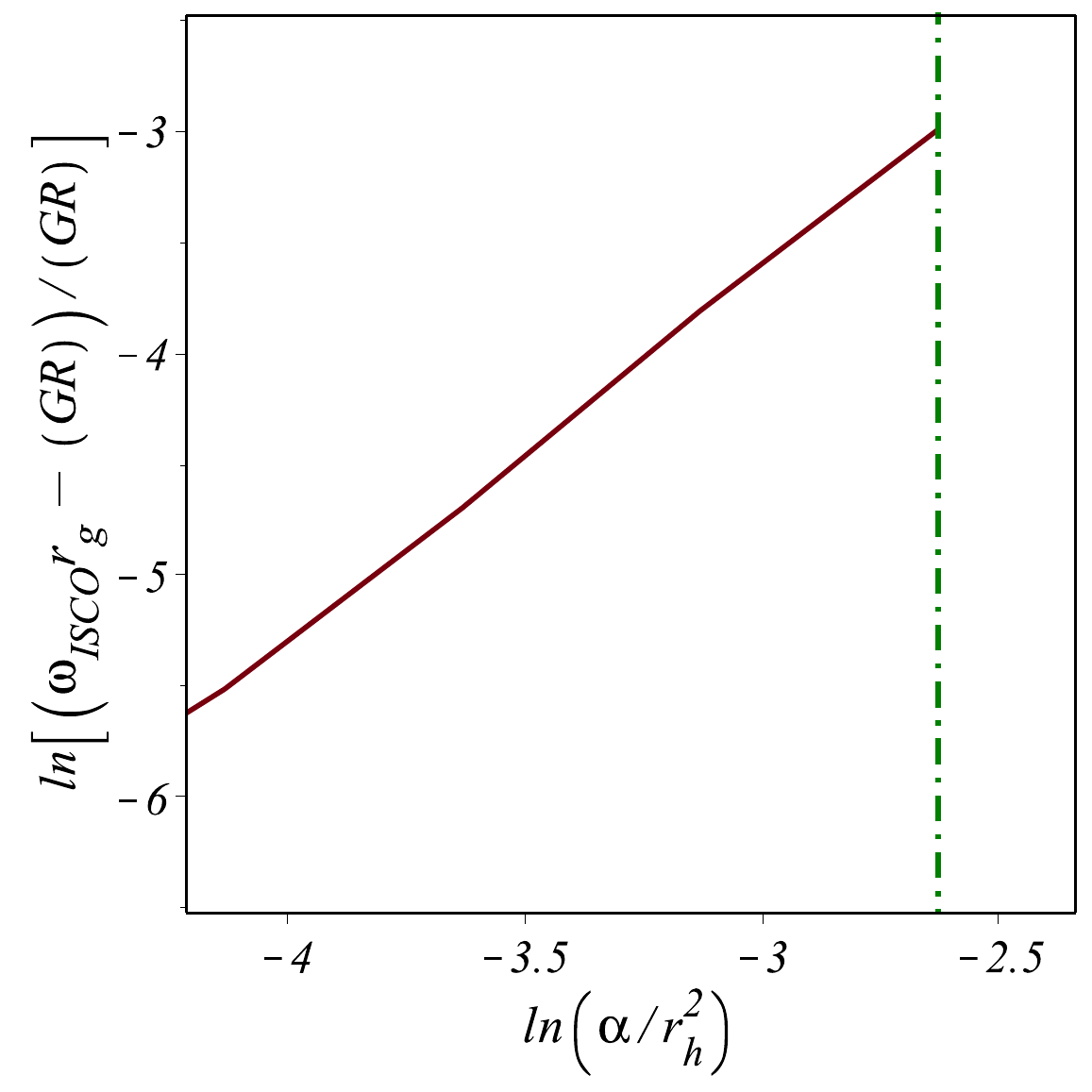}
\caption{Fractional deviation of the ISCO angular frequency times the gravitational radius, $\omega_{ISCO}\,r_g$, from the value it would have in GR, denoted as (GR). The green, dot-dashed vertical line is the  largest value  $\alpha/r_h^2$  can take without the solution having a naked singularity.} 
\label{fig:wISCOrg}
\end{figure}

\begin{figure}
\includegraphics[height=2.7in,width=2.7in]{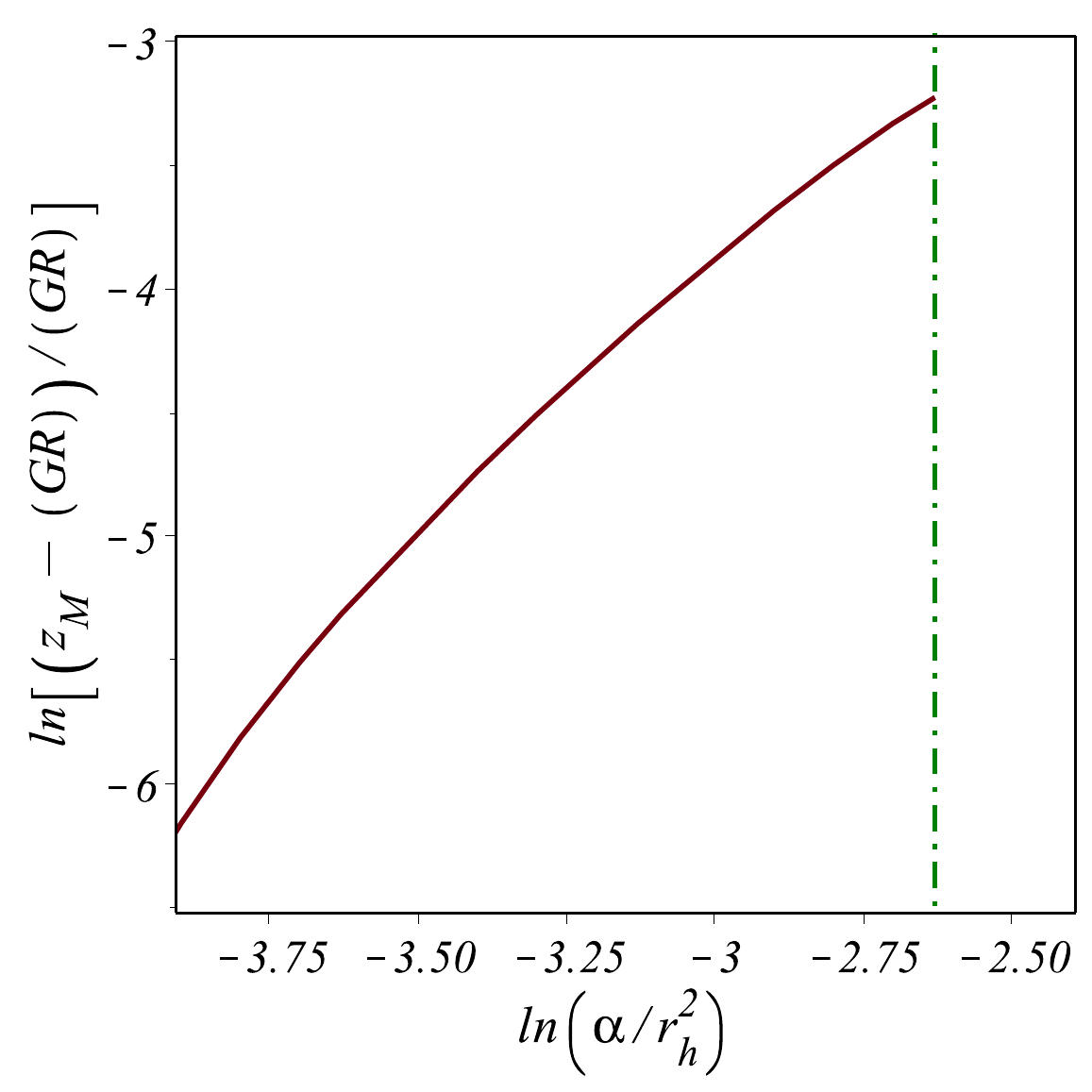}
\caption{Fractional deviation of the maximum redshift of a photo emitted from the ISCO, $z_M$, from the value it would have in GR, denoted as (GR). The green, dot-dashed vertical line is the  largest value  $\alpha/r_h^2$  can take without the solution having a naked singularity.} 
\label{fig:zM}
\end{figure}

\begin{figure}
\includegraphics[height=2.7in,width=2.7in]{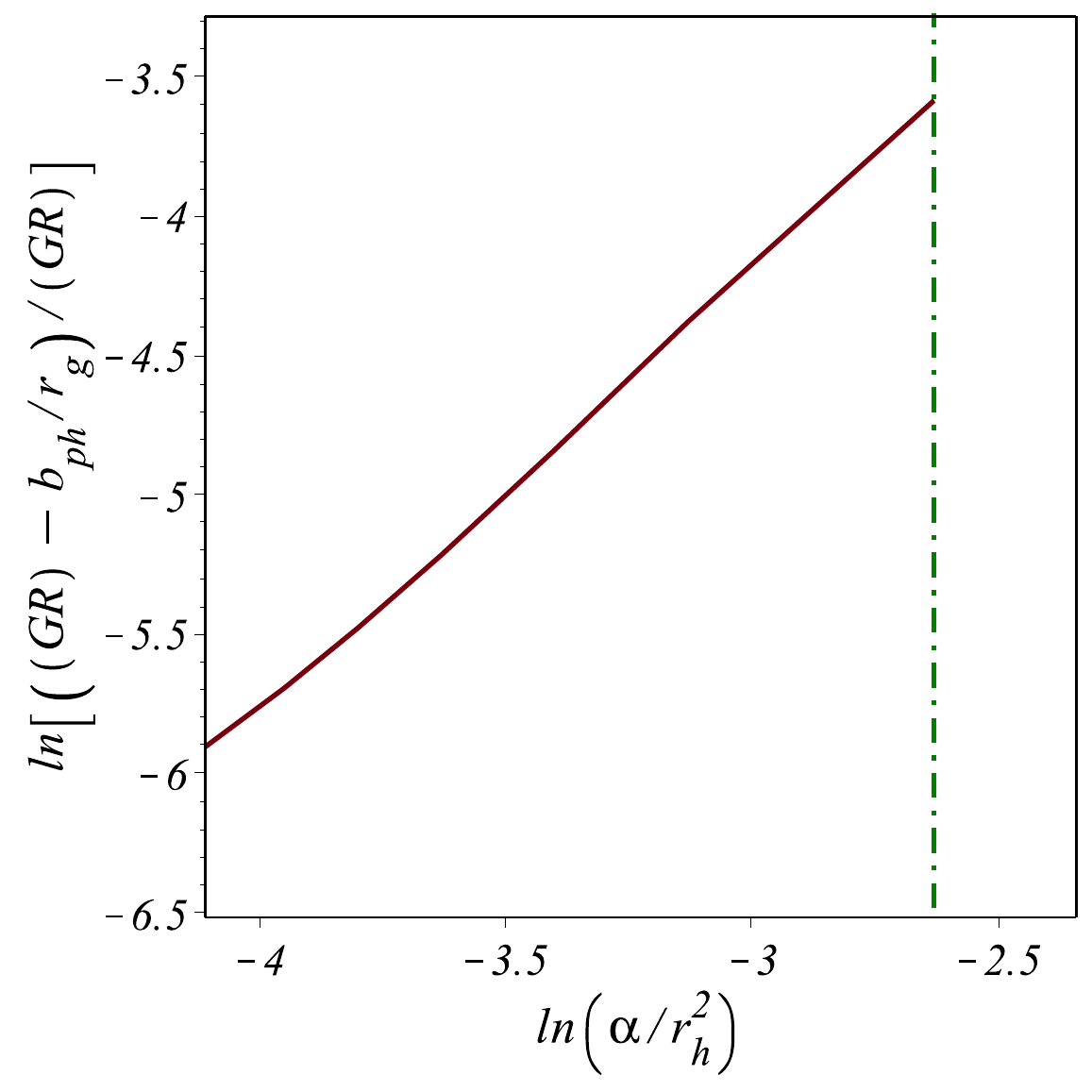}
\caption{Fractional deviation of the impact parameter of the photon circular orbit over the gravitational radius, $b_{ph}/r_g$, from the value it would have in GR, denoted as (GR). The green, dot-dashed vertical line is the  largest value  $\alpha/r_h^2$  can take without the solution having a naked singularity.} 
\label{fig:bphrg}
\end{figure}

\begin{figure}
\includegraphics[height=2.7in,width=2.7in]{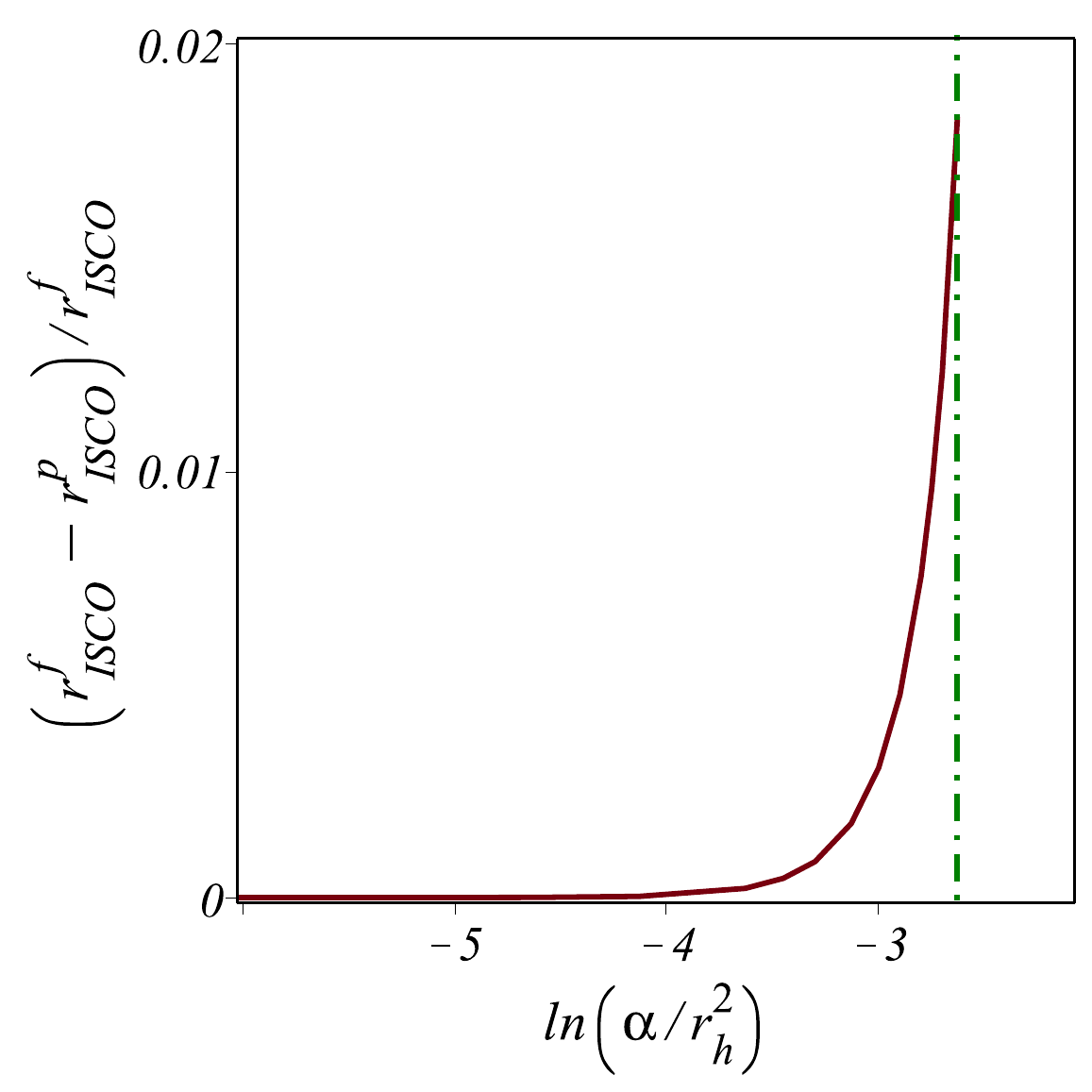}
\caption{$r^f_{ISCO}$ and $r^p_{ISCO}$ are the ISCO radii for the numerical and perturbative solutions respectively. The green, dot-dashed vertical line is the  largest value  $\alpha/r_h^2$  can take without the solution having a naked singularity.} 
\label{fig:FPrISCO}
\end{figure}

It is worth pointing out that in the latter 3 figures, those for $\omega_{ISCO}r_g$, $z_M$ and $b_{ph}/r_g$, we do not consider values of $\alpha$ as low as in the figures for $r_g$ and $r_{ISCO}$. This is because in order to compute $\omega_{ISCO}r_g$, $z_M$ and $b_{ph}/r_g$ one needs to numerically solve Eq.~(\ref{iscorad}) or Eq.~(\ref{phrad}) as an intermediate step, which intrinsically reduces the numerical accuracy. Going to smaller values of $a/r_h^2$ would require more accuracy in the numerical implementation, but that seems unnecessary, given that the deviation from GR is already very small for the values of $\alpha/r_h^2$ computed and for smaller $\alpha/r_h^2$ one can safely use the perturbative solution we have obtained in Section \ref{sec:hasi}.

The exact, numerical solution has been used to produce all of the plots. One may wonder how well the perturbative solution approximates the full numerical one. Fig.~\ref{fig:Gcompare} already suggests that the $\mc{O}(\alpha^2)$ approximation is relatively good sufficiently far from the finite area singularity. The presence of the latter appears to be a highly nonlinear effect, so one expects  the perturbative solution to become less and less accurate as the finite radius singularity of the black hole approaches the horizon, i.e., as $\alpha$ approaches $r_h^2/8\sqrt{3}$. In practice, we find that even when $\alpha$ is only slightly less than $r_h^2/8\sqrt{3}$, the $\mc{O}(\alpha^2)$ perturbative solution at most deviates from the full solution by a few percents for the quantities we consider. See e.g.~Fig.~\ref{fig:FPrISCO} for a comparison between the ISCO radius of the numerical and the perturbative solution. If we were to plot the corresponding quantities computed from the perturbative solution in Figs.~\ref{fig:rg}-\ref{fig:bphrg}, the perturbative solution line would overlap with the full solution line until when $\alpha$ is very close to $r_h^2/8\sqrt{3}$.

\section{Conclusions}
\label{sec:concl}

We have investigated the existence of hairy black holes in shift-symmetric generalized galileon (SSGG), which is the most general scalar-tensor theory that has shift symmetry for the scalar and whose equations of motion have up to second order derivatives. A  no-hair theorem for static, spherically symmetric black holes in  this theory  has been put forward in Ref.~\cite{Hui:2012qt}. In Ref.~\cite{Sotiriou:2013qea} we  showed that one assumption in the last step of the proof does not generically hold in SSGG. We have elaborated on this point further and argued in detail that there are 2 cases in which the assumption does not hold:  1) models of SSGG that are Lorentz-violating; 2) and theories that are fully Lorentz invariant and where a linear scalar-Gauss-Bonnet coupling $\phi \mc{G}$ is present in the Lagrangian. From the effective field theory point of view, it seems unnatural to exclude this scalar-Gauss-Bonnet operator, as it is not forbidden by the underlying symmetries of the theory. So in a sense existence of hairy black holes seems to be a generic feature of SSGG.

The evasion of the no-hair arguments does not necessarily imply the actual existence of hairy black holes. To show that hairy black holes do indeed exist  we have presented here explicit numerical solutions in the theory of (\ref{gbaction}), the simplest in the evading Lagrangian containing the term (\ref{genevading}). These solutions features a finite surface singularity, which is cloaked by the event horizon if the coupling constant $\alpha$ is sufficiently small. Black holes of this type have a minimum size for a give coupling  $\alpha$. The scalar hair is secondary, according to the terminology of \cite{Coleman:1991ku}, {\em i.e.}~the scalar charge is not an independent parameter, but it is instead fixed in terms of the mass of the black hole. This last condition is imposed by the requirement that the scalar be regular on the horizon. Note that the fact that the scalar hair is secondary does not necessarily make them any less noteworthy. 
In our case, since the mass of the black hole and the scalar hair are related, changes in the scalar configuration can affect the black hole mass and vice versa. 

We have also constructed analytic  solutions perturbatively up to second order in the small $\alpha$ limit. The advantage of these solutions is that they are explicit and this make their properties more obvious. Remarkably, at first order in the coupling the spacetime does not deviate from the Schwarzschild solution and one simply has a non-trivial configuration of the scalar field in this geometry. The scalar charge is already fixed in terms of the mass at this order. This solution might be considered the only solution one can trust if the action is considered a product of some truncation, with $\alpha$ being the expansion parameter, as in this case $\mc{O}(\alpha^2)$ corrections in the action and field equation would have been anyway neglected. The perturbative solution to $\mc{O}(\alpha^2)$ deviates from the Schwarzschild metric. The scalar field profile instead does not receive any $\mc{O}(\alpha^2)$ corrections. To this order, the perturbative solution fails to capture the presence of a finite area singularity. In fact, we have shown that the perturbative approximation breaks down in the vicinity of the finite area singularity radius. Hence, one can conclude that non-linear effects associated with the $\phi {\mc G}$ term should be crucial for the formation of this singularity.
On the other hand, in the regime where the small coupling perturbative treatment is valid, the perturbative solution seems to be an excellent approximation of the full numerical solution. 

Finally, we have  investigated the phenomenological properties of  the black holes we have generated. We have computed several observables for different values of the coupling $\alpha$. If we require the finite radius singularity be well within the horizon, the deviations from the Schwarzschild black hole are negligibly small. Hence, it would be rather challenging to detect the presence of a linear scalar-Gauss--Bonnet coupling using black hole observations, even with upcoming astronomical observations. However, any result in this direction should be considered highly preliminary for two reasons: firstly, astrophysical black holes are rotating and our black hole do not capture this feature. Certain effects might be much more sensitive to rotation. Additionally, perturbation of these black holes will involve an extra scalar excitation, which could lead to detectable deviation from GR. (Examples that support both claims can be found in Refs.~\cite{Barausse:2008xv,Cardoso:2011xi,Cardoso:2013opa,Cardoso:2013fwa}).

\begin{acknowledgments}
We would like to thank Enrico Barausse, Arif Mohd, Paolo Pani and Ian Vega for helpful discussions. The research leading to these results has received funding from the European Research Council under the European Union's Seventh Framework Programme (FP7/2007-2013)/ERC Grant Agreement n.~306425 ``Challenging General Relativity''.
\end{acknowledgments}

\appendix

\section{The most general shift-symmetric generalized galileon}
\label{sec:ssgg}

In this section, we will identify the most general shift symmetric subset of generalized galileons \cite{Deffayet:2011gz} (or Horndenski's theory \cite{Horndeski:1974wa}). The Lagrangian for generalized galileons can be written as
\bal
\label{gglt}
\mc{L} &= \mc{L}_2+\mc{L}_3+\mc{L}_4+\mc{L}_5     ,
\\
\mc{L}_2 &= K(\phi,X)     ,
\\ 
\mc{L}_3 &= -G_3(\phi,X) \Box \phi     ,
\\
\mc{L}_4 &= G_4(\phi,X) R + G_{4X} \left[ (\Box \phi)^2 -(\nabla_\mu\nabla_\nu\phi)^2 \right]     ,
\\
\mc{L}_5 &= G_5(\phi,X) G_{\mu\nu}\nabla^\mu \nabla^\nu \phi - \frac{1}{6}G_{5X} \bigg[ (\Box \phi)^3 
\nn
&~~~~ - 3\Box \phi(\nabla_\mu\nabla_\nu\phi)^2 + 2(\nabla_\mu\nabla_\nu\phi)^3 \bigg]     ,
\eal
where $K, G_3, G_4, G_5$ are arbitrary functions of $\phi$ and $X$ and here $f_X$ stands for $\pd f(\phi,X)/\pd X$. The full equations of motion for this theory have been derived in \cite{Kobayashi:2011nu}  and independently in \cite{Gao:2011mz}. They are rather long and cumbersome, so we will not displayed them fully. Our proof does rely on the explicit form of the metric equation of motion, the part that is proportional to $g_{\mu\nu}$ to be precise, as we discuss below.

To get shift-symmetric generalized galileon (SSGG), we require the equations of motion be invariant under  $\phi \to \phi+\epsilon$. That is, we require $\delta \mc{E}_{\mu\nu} = \epsilon \pd \mc{E}_{\mu\nu}/\pd \phi$ to vanish identically, where $\mc{E}_{\mu\nu}=0$ is the field equation for $g_{\mu\nu}$. Note that requiring an expression to vanish identically is different from solving an equation to obtain solutions; we should require the expression to vanish for any field configurations. This means, in our case, terms with the same derivative and curvature tensor structure should cancel exactly, which therefore imposes constraints on the arbitrary functions $G_i(\phi,X)$. (One might be interested in imposing shift symmetry at the level of the Lagrangian and require terms with the same structure to cancel out, but in that case the ``same structure'' becomes ambiguous, as one is allowed to integrate by parts.)

Let us now focus on the part of $\delta \mc{E}_{\mu\nu}$ that is proportional to $g_{\mu\nu}$, which is readily available in \cite{Kobayashi:2011nu}. First, we notice that there is a term proportional to $-g_{\mu\nu}G_{5X\phi} \nd^{[\rho}\nd_\rho\phi\nd^{\sigma}\nd_\sigma\phi\nd^{\lambda]}\nd_\lambda\phi$, which can not be canceled by any other term, so we have to set 
\be
G_{5X\phi}=0
\ee
in $\delta \mc{E}_{\mu\nu}$. Then, the $g_{\mu\nu}$ part of $\delta \mc{E}_{\mu\nu}$ becomes
\bwt
\bal
\delta \mc{E}^{(2)}_{\mu\nu} & \supset  \epsilon  g_{\mu\nu} \left( -\frac12 K_\phi \right)     ,
\\
\delta \mc{E}^{(3)}_{\mu\nu} & \supset  \epsilon  g_{\mu\nu} 
\left( G_{3\phi\phi}X +\frac12 G_{3\phi X} \nd_\lambda \nd_\rho \phi \nd^\rho\phi\nd^\lambda\phi \right)     ,
\\
\delta \mc{E}^{(4)}_{\mu\nu} & \supset  \epsilon  g_{\mu\nu} 
\Big( 
G_{4\phi\phi} \Box \phi -2X G_{4\phi\phi\phi}  -2G_{4\phi\phi X} (\nd_\rho\nd_\sigma\phi)^2
- 2 G_{4X\phi\phi}X\Box\phi 
\nn
&\qquad~~~~ -G_{4XX\phi}\nd_\lambda \nd^\rho \phi \nd^\lambda \phi \nd_\rho\phi\Box\phi   
+G_{4XX\phi} \nd_\rho \nd_\lambda\phi \nd_\sigma\nd^\lambda\phi \nd^\rho\phi \nd^\sigma\phi
\nn
&\qquad~~~~
+G_{4\phi X} \nd^{[\rho}\nd_\rho\phi\nd^{\sigma]}\nd_\sigma\phi  
- G_{4\phi X}R^{\rho\sigma}\nd_\rho\phi\nd_\sigma\phi 
\Big)     ,
\\
\delta \mc{E}^{(5)}_{\mu\nu} |_{G_{5\phi X}=0} & \supset  \epsilon  g_{\mu\nu} 
\left(\phantom{\frac11}\!\!\!\!
G_{5\phi\phi}R^{\rho\sigma} \nd_\rho\phi\nd_\sigma\phi 
- G_{5\phi\phi}\nd^{[\rho}\nd_\rho\phi\nd^{\sigma]}\nd_\sigma\phi
+G_{5\phi\phi\phi}X\Box\phi \right.
\nn
&\qquad~~~~ \left. +\frac12 G_{5\phi\phi\phi} \nd_\rho\phi\nd_\sigma\phi\nd^\rho\nd^\sigma\phi
\right)     ,
\eal
\ewt
where the superscript ${}^{(i)}$ refers to the relevant part of the Lagrangian from which the term originates.
Requiring terms with the same derivative and curvature structure to cancel, we get the following differential equations for the four arbitrary functions:
\bal
G_{4\phi X} - G_{5\phi\phi} &= 0     ,
\\
-2G_{4\phi\phi X} + \frac12 G_{5\phi\phi\phi} &=0     ,
\\
G_{4\phi\phi} -2G_{4\phi\phi X}X + G_{5\phi\phi\phi}X &=0     ,
\\
G_{4\phi X X} & =0      ,
\\
G_{3\phi X } & =0      ,
\\
-\frac12 K_\phi +G_{3\phi\phi}X &=0     .
\eal
The general solution of these equations is:
\bal
G_5&= c_1 \phi^2+c_2\phi + G_5(X)     ,
\\
G_4&=2c_1\phi X + G_4(X)     ,
\\
G_3&=f_3(\phi) + c_3 \phi +G_3(X)     ,
\\
K  &=  2Xf_{3\phi}(\phi)  + K(X)     ,
\eal
where $c_1$, $c_2$ and $c_3$ are constants, $f_3(\phi)$ is an arbitrary function of $\phi$, and we have abused the notation slightly to introduce new arbitrary functions of $X$. Note that, unlike flat space galileon theory, the scalar tadpole term in a covariant theory is not shift symmetric, as a shift in the scalar tadpole term simply changes the value of the cosmological constant.

The reason why $c_3 \phi$ and $f_3(\phi)$ can be present without compromising shift symmetry is because there is degeneracy between $\mc{L}_2$ and $\mc{L}_3$ in choosing Lagrangian terms that gives rise to the same dynamics. The same is true for $c_1$ and $c_2$ in the Lagrangian $\mc{L}_4$ and $\mc{L}_5$. More specifically,  the following identities hold
\bal
&~~~~f_3(\phi) \Box \phi 
\nn
&= 2Xf_{3\phi}(\phi)  + {\rm total~derivative}     ,
\\
 &~~~~ X R + (\Box\phi)^2 - (\nd_\mu\nd_\nu \phi)^2 
 \nn
 &= -\phi G_{\mu\nu} \nd^{\mu}\nd^{\nu}\phi + {\rm total~derivative}     ,
\\
& ~~~~\phi X R +\phi[ (\Box\phi)^2 - (\nd_\mu\nd_\nu \phi)^2] 
\nn
&= -\frac12 \phi^2 G_{\mu\nu} \nd^{\mu}\nd^{\nu} \phi+ {\rm total~derivative}     .
\eal
The first two of these relations have been pointed out in \cite{Kobayashi:2011nu}. Making use of the above relations, after some cancellations, the Lagrangian terms with $c_1$ and $f_3(\phi)$ are only total derivatives, so they play no role in determining the dynamics of the system, so we can set $c_1=0$ and $f_3(\phi)=0$ without lost of generality. The Lagrangian terms with $c_2$ and $c_3$ can be absorbed into redefined $G_4(X)$ and $K(X)$, so we can also set $c_2=0$ and $c_3=0$. Therefore, only the four arbitrary functions $K(X), G_3(X),G_4(X), G_5(X)$ are left. 

Lastly, we have only consider a fraction of the full equations of motion, but a transformation is a symmetry only if all the equations of motion are invariant under it. However, at this point, if we check the equations of motion or simply the Lagrangian, the theory is already manifestly shift symmetric. 

In summary, the most general shift-symmetric generalized galileon is given by Lagrangian \label{gglt} with the four arbitrary functions replaced by
\bal
G_5(\phi,X) &\to G_5(X)     ,
\\
G_4(\phi,X) &\to G_4(X)     ,
\\
G_3 (\phi,X) &\to G_3(X)     ,
\\
K (\phi,X) &\to K(X)     .
\eal

\section{Gauss-Bonnet term as a total derivative}
\label{sec:GBterm}

It is most convenient to formulate the Gauss-Bonnet term as a total derivative in the vielbein formalism $e^A_\mu$. First, note that the curvature form $\Omega^{A}{}_B$ can be written as
\be
\Omega^{A}{}_B=\uD\omega^{A}{}_B=\ud\omega^{A}{}_{B}+\omega^{A}{}_C\wedge\omega^{C}{}_{B}     ,
\ee
where $\uD$ is the covariant exterior derivative and $\omega^A{}_B$ is the spin connection satisfying $\ud e^A + \omega^A{}_B\wedge e^B=0$, and the second Bianchi identity is 
\be
\uD\Omega^{AB} = \ud \Omega^{AB} + \omega^A{}_C \wedge \Omega^{CB} + \omega^B{}_C \wedge \Omega^{AC} =0   .
\ee
With these relations and the form of the Gauss-Bonnet term in terms of $\Omega^{A}{}_B$, we have 
\bal
S &= \int \ud^4 x \sqrt{-g} \mc{G}  = \int \ud^4 x  \sqrt{-g} \nd_\mu \tilde{\mc{G}}^\mu
\\
&= \int \Omega^{AB}\wedge \Omega^{CD}\epsilon_{ABCD}
\\
&= \int \ud(\omega^{A}{}_{B}\wedge \Omega^{C}{}_{D}\epsilon_A{}^B{}_C{}^D)
\\
&=\frac12 \int \pd_\mu (\epsilon_A{}^B{}_C{}^D \omega_\nu^{A}{}_{B} R_{\rs}{}^{C}{}_{D}) 
\nn
&~~~~~~~~~~~~~~~~~~\cdot\ud x^\mu \wedge \ud x^\nu \wedge \ud x^\rho \wedge \ud x^\sigma
\\
&= -\frac12 \int \ud^4 x  \pd_\mu (\sqrt{-g}\epsilon^{\mn\rs}\epsilon_A{}^B{}_C{}^D \omega_\nu^{A}{}_{B} R_{\rs}{}^{C}{}_{D}) 
\\
&=  \int \ud^4 x  \sqrt{-g} \nd_\mu \left(\frac12\epsilon^{\mn\rs}\epsilon_{\alpha\beta}{}^{\lambda\eta} \omega_\nu^{\alpha}{}_{\lambda} R{}^{\beta}{}_{\eta}{}_{\rs}\right)       ,
\eal
where we have defined $\epsilon_{\mn\rs}|_{\mn\rs=0123}=\sqrt{-g}$ and $\epsilon_{ABCD}|_{ABCD=0123}=1$. So the quantity $\tilde{\mc{G}}^\mu$ introduced in Section \ref{sec:countexam} is given by
\be
\tilde{\mc{G}}^\mu = \frac12\epsilon^{\mn\rs}\epsilon_{\alpha\beta}{}^{\lambda\eta} \omega_\nu^{\alpha}{}_{\lambda} R{}^{\beta}{}_{\eta}{}_{\rs}   .
\ee

\section{Variation of the scalar-Gauss-Bonnet term}
\label{sec:sGBt}

Here we derive the contribution of the action term
\bal
S_{\phi \rm GB}&=\int\ud^4x\sqrt{-g} \phi \mc{G} 
\nn
&=\int\ud^4x\frac14 \sqrt{-g}\phi\delta^{\mn\rs}_{\alpha\beta\gamma\delta} R_{\mn}{}^{\alpha\beta} R_{\rs}{}^{\gamma\delta}      ,
\eal
 to the field equation of the metric, where $\delta^{\mn\rs}_{\alpha\beta\gamma\delta} = 4! \delta^\mu_{[\alpha} \delta^\nu_\beta \delta^\rho_\gamma \delta^\sigma_{\delta]} = -\epsilon^{\mn\rs}\epsilon_{\alpha\beta\gamma\delta}$. $\epsilon^{\mn\rs}$ is the Levi-Civita tensor. In the field equations for the metric $\delta S_{\phi \rm GB}/\delta g^{\mu\nu}$, there are terms containing $\pd\phi$, as well as terms containing $\phi$ {\it a priori}. According to the Gauss-Bonnet theorem, the terms containing $\phi$ should cancel each other (otherwise the metric equation of motion is not invariant under $\phi\to \phi+c$),  so we only need to keep track of terms containing $\pd\phi$ when varying $S_{\phi \rm GB}$. We will use the equality ``$\dot{=}$'' when these $\phi$ terms are dropped. Also, the variation in the following is only with respect to the metric.

Making use of $\dd R^\rho{}_{\sigma\mn} = 2 \nd_{[\mu}\dd \Gamma^\rho_{\nu]\sigma}$ and $\dd \Gamma^\rho_{\mn} = \frac12 g^\rs [ \nd_\mu \dd g_{\nu\sigma} + \nd_\nu \dd g_{\mu\sigma} - \nd_\sigma \dd g_{\mn}]$, we have 
\bal
\delta S_{\phi \rm GB} & \dot{=}  \int\ud^4x \frac12\sqrt{-g}\phi\delta^{\mn\rs}_{\alpha\beta\gamma\delta} R_{\mn}{}^{\alpha\beta} \dd R_{\rs}{}^{\gamma\delta} 
\nn
&= \int\ud^4x (\dd \mc{T}_1 +\dd \mc{T}_2 + \dd \mc{T}_3)     ,
\eal
 where
 \bal
 \dd \mc{T}_1 & = \frac12\sqrt{-g}\phi\delta^{\mn\rs}_{\alpha\beta\gamma\delta} R_{\mn}{}^{\alpha\beta} \nd^\gamma \nd^\delta \dd g_{\rs} = 0     ,
 \\
  \dd \mc{T}_2 & =\frac12\sqrt{-g}\phi\delta^{\mn\rs}_{\alpha\beta\gamma\delta} R_{\mn}{}^{\alpha\beta} \nd^\gamma \nd_\sigma  (g^{\delta\lambda} \dd g_{\lambda\rho} )     ,
 \\
  \dd \mc{T}_3 &=  \dd \mc{T}_2 
  \nn
  & =  - \frac12\sqrt{-g}\phi\delta^{\mn\rs}_{\alpha\beta\gamma\delta} R_{\mn}{}^{\alpha\beta} \nd^\gamma \nd_\rho  (g^{\delta\lambda} \dd g_{\lambda\sigma} )     .
 \eal
$\dd \mc{T}_1$ vanishes because $\rho$ and $\sigma$ are antisymmetrized. So, by partial integrations, we have
\bal
&~~~~\delta S_{\phi \rm GB} 
\nn
& \dot{=}  -\int\ud^4x  \sqrt{-g}\phi\delta^{\mn\rs}_{\alpha\beta\gamma\delta} R_{\mn}{}^{\alpha\beta} \nd^\gamma \nd_\sigma  (g_{\rho\lambda} \dd g^{\delta\lambda} )
\\
& \dot{=} \int\ud^4x  \sqrt{-g}    \nd_\sigma\left(\nd^\gamma \phi \:\epsilon^{\mn\rs}\epsilon_{\alpha\beta\gamma\delta} R_{\mn}{}^{\alpha\beta}   \right) g_{\rho\lambda} \dd g^{\delta\lambda} 
\\
& {=} \int\ud^4x \; \frac12 \sqrt{-g} (g_{\rho\mu}g_{\delta\nu}  + g_{\rho\nu}g_{\delta\mu})  \nn
&~~~~~~~~~~~~~\cdot  \nd_\sigma\left(\pd_\gamma \phi \:\epsilon^{\lambda\eta\rs}\epsilon^{\alpha\beta\gamma\delta} R_{\lambda\eta\alpha\beta}   \right) \dd g^{\mu\nu}    .
\eal
Note that $\epsilon^{\gamma\delta\alpha\beta}\epsilon^{\rs\lambda\eta} R_{\lambda\eta\alpha\beta}/4$ is the double dual Einstein tensor and divergence-free.

\subsection{Equations of motion: spherical symmetry}
\label{sec:eomss}

Here we list the relevant equations of motion terms in spherical symmetry:
\bal
G^t{}_t&=-\frac{1}{r^2} +\frac{\eta  }{r^2 e^B}- \frac{\eta B'}{re^B}     ,
\\
G^r{}_r&=-\frac{1}{r^2} +\frac{\eta  }{r^2 e^B}+ \frac{\eta A'}{re^B}     ,
\\
G^\theta{}_\theta&=G^\varphi{}_\varphi
\nn
&=\eta\left(\frac{A'-B'}{2re^B}+\frac{A'(A'-B')}{4e^B} +\frac{A''}{2e^B}\right)     ,
\\
\mc{T}^t{}_t &= \frac{4\alpha(\eta e^B-3)\phi'B'}{r^2 e^{2B}}  -\frac{\eta\phi'^2}{2e^B}
\nn
&~~~~ -\frac{8\alpha(\eta  e^B-1)\phi''}{r^2e^{2B}}     ,
\\
\mc{T}^r{}_r &= \frac{12\alpha\phi'A'}{r^2 e^{2B}} -\frac{4\eta\alpha\phi'A'}{r^2e^B} +\frac{\eta\phi'^2}{2e^B}     ,
\\
\mc{T}^\theta{}_\theta &=\mc{T}^\varphi{}_\varphi 
\nn
&=
-\frac{\eta\phi'^2}{2e^B} +\frac{2\alpha (A'-3B')  A'\phi'}{re^{2B}} 
\nn
&~~~~ +\frac{4\alpha (\phi'A''+\phi''A')}{re^{2B}}     ,
\\
\Box \phi&= \frac{2\eta\phi'}{re^B} +\frac{\eta(A'-B')\phi'}{2e^B} +\frac{\eta \phi''}{e^B}     ,
\\
\mc{G}&=\frac{2(1-\eta e^B)A'^2 +2(\eta e^B-3)A'B'  }{r^2e^{2B}}  
\nn
&~~~~ +\frac{ 4(1-\eta e^B)A''}{r^2e^{2B}}  .
\eal

\section{No-hair for slowly rotating black holes}  
\label{sec:nhsrbh}

Here we present a short argument which suggests that, for a gravity model containing the metric and a number of scalars, a no-hair theorem for spherical symmetric black holes can be readily generalized to the slowly rotating black holes. That is, for a multi-scalar-tensor theory, if a static, spherically symmetric black hole does not have scalar hair, then its slowly rotating counterpart will not have hair either. This argument has been already put forth in our Letter \cite{Sotiriou:2013qea}.

First, we note that the most general stationary axisymmetric metric to first order in rotation is given by \cite{hartle67}
\bal
\ud s^2 &= -H(r)\ud t^2 +Q(r)  \ud r^2 + r^2  [\ud \theta^2 +\sin^2\theta \ud \varphi^2] 
\nn
&~~~~~ - 2\epsilon \omega(r,\theta)r^2\sin^2\theta  \ud t \ud \varphi + \mathcal{O}(\epsilon^2)   ,
\eal
where $H(r)$ and $Q(r)$ correspond to the static, spherically symmetric seed solution, $\omega(r,\theta)$ parametrizes the correction induces by the rotation, and $\epsilon$ is just a book-keeping parameter for the expansion.  On the other hand, the $n$-th scalar field with axisymmetry can be generally expanded as
\be
\phi^{n}(x) = \phi^n_0(r) +\phi^n_1(t,r,\theta) \epsilon + \mathcal{O}(\epsilon^2)   .
\ee
Now, due to axisymmetry, the system is invariant under the combined operation of $\omega(r,\theta)\to -\omega(r,\theta)$ and $\varphi \to -\varphi$. The metric is clearly invariant under this  operation.  However, $\phi^n_1(t,r,\theta)$ is by definition linear in the rotation so it should change sign every time the direction of rotation is reversed. Hence, for the scalar fields $\phi^n$ to be invariant under $\omega(r,\theta)\to -\omega(r,\theta)$ and $\varphi \to -\varphi$, we must have
\be
\phi^n_1(t,r,\theta) = 0  \Longrightarrow  \phi^n(x) = \phi^n_0(r)+\mathcal{O}(\epsilon^2)   .
\ee
Thus, the scalar fields $\phi^n(x)$ do not acquire a correction at first order in rotation.

\end{document}